\documentstyle[12pt]{article}
\begin{document}
\thispagestyle{empty}
\def\cqkern#1#2#3{\copy255 \kern-#1\wd255 \vrule height #2\ht255 depth
   #3\ht255 \kern#1\wd255}
\def\cqchoice#1#2#3#4{\mathchoice%
   {\setbox255\hbox{$\rm\displaystyle #1$}\cqkern{#2}{#3}{#4}}%
   {\setbox255\hbox{$\rm\textstyle #1$}\cqkern{#2}{#3}{#4}}%
   {\setbox255\hbox{$\rm\scriptstyle #1$}\cqkern{#2}{#3}{#4}}%
   {\setbox255\hbox{$\rm\scriptscriptstyle #1$}\cqkern{#2}{#3}{#4}}}
\def\CC{\mathord{\cqchoice{C}{0.65}{0.95}{-0.1}}}
\def\x{\stackrel{\otimes}{,}}
\def\y{\stackrel{\circ}{\scriptstyle\circ}}
\def\proof{\noindent Proof. \hfill \break}
\def\a{\begin{eqnarray}}
\def\b{\end{eqnarray}}
\def\p{{1\over{2\pi i}}}
\def\Q{{\scriptstyle Q}}
\def\P{{\scriptstyle P}}
\renewcommand{\thefootnote}{\fnsymbol{footnote}}

\newpage
\setcounter{page}{0}
\pagestyle{empty}

\centerline{\LARGE On Matrix KP and Super-KP Hierarchies}
\centerline{\LARGE in the Homogeneous Grading.}
\vspace{1truecm} \vskip0.5cm

\centerline{\large F. Toppan}
\vskip.5cm
\centerline{Dipartimento di Fisica}
\centerline{Universit\`{a} di Padova}
\centerline{Via Marzolo 8, I-35131 Padova}
\centerline{\em E-Mail: toppan@mvxpd5.pd.infn.it}
\vskip1.5cm
\centerline{\bf Abstract}
\vskip.5cm
Constrained KP and super-KP hierarchies of integrable equations
(generalized NLS hierarchies) are systematically produced through a
Lie algebraic AKS-matrix framework associated to the homogeneous grading.
The role played by different regular elements to define the
corresponding hierarchies is analyzed as well as
the symmetry properties under the Weyl group transformations.
The coset structure of higher order hamiltonian densities is proven.\par
For a generic Lie algebra
the hierarchies here considered are integrable and essentially
dependent on continuous free parameters.
The bosonic hierarchies
studied in \cite{{FK},{AGZ}} are obtained as
special limit restrictions on hermitian symmetric-spaces.\par
In the supersymmetric case the homogeneous grading is introduced
consistently by using alternating sums of bosons
and fermions in the spectral parameter power series.\par
The bosonic hierarchies obtained
from ${\hat {sl(3)}}$ and the supersymmetric ones derived
from the $N=1$ affinization of $sl(2)$,
$sl(3)$ and $osp(1|2)$ are explicitly constructed. \par
An unexpected result is found: only a restricted subclass of the
$sl(3)$ bosonic hierarchies can be supersymmetrically extended
while preserving integrability.
\vfill
\rightline{DFPD 95/TH/31}
\rightline{May 1995}
\newpage
\pagestyle{plain}
\renewcommand{\thefootnote}{\arabic{footnote}}
\setcounter{footnote}{0}

\section{Introduction.}

\indent

Recently a lot of research has been devoted
to the hierarchies of integrable differential equations due to their
connection with the discretized version (matrix-model formulation) of
the $2$-dimensional gravity (see e.g. \cite{DiF} for a review). It is
in fact by now clear
that constrained KP flows \cite{{bonora},{arat}} define the partition
functions of single and multi-matrix models.\par
Moreover it has been
suggested \cite{alv} that supersymmetric hierarchies describe
the $2$-dimensional supergravity, even if no matrix-model formulation
is at present available in this case.\par
The problem of classifying all possible (both bosonic and
supersymmetric) hierarchies is therefore quite a crucial one.\par
In the bosonic case a strategy, based on a
generalized Drinfeld-Sokolov approach, has been developed in many
papers \cite{{DS},{genkdv},{feh}}.
Basically to produce integrable hierarchies
the following ingredients are needed: a
matrix-type Lax operator valued on a Lie algebra ${\cal G}$; a
suitable ${\bf Z}$-grading for ${\cal G}$ and the existence of
constant non-zero graded regular elements for the algebra (for details
see \cite{feh}). The problem of classifying hierarchies is therefore
reduced to the Lie-algebraic problem of determining the
acceptable gradings
and the corresponding regular elements. This problem has been solved
for non-exceptional Lie algebras as well as for (at least some of) the
exceptional ones.
Therefore in the bosonic case the situation seems
completely satisfactory apart perhaps some questions like e.g.
do different
Lie algebras and different gradings always produce different
hierarchies? Which is the role played by different regular elements once
chosen a given Lie algebra and a given grading? This second question will
be addressed in this paper for the special case of the homogeneous grading
and it will be shown that indeed different regular elements induce different
hierarchies; moreover, in the general case, the integrability is
preserved even in presence of an essential dependence on continuous
free parameters. \par
In the supersymmetric case the situation is much less satisfactory:
supersymmetric versions of matrix super-KP hierarchies \cite{IK}
have been constructed only for Lax operators which take values on
superalgebras and are associated to the principal grading (generalized
super-KdV hierarchies); moreover, since the Lax operator is in this
case a fermionic object, in its turn
the constant regular element must be fermionic;
as a consequence only superalgebras which admit a presentation
in terms of fermionic simple roots can produce
super-hierarchies \cite{IK}.\par
On the other hand integrable super-hierarchies which do not fit in the
above scheme have actually been constructed (see
\cite{{roe},{brun},{toppan}}).
The lack of a clear Lie algebraic understanding of such hierarchies
makes difficult to find their generalizations; moreover the
recognition of their integrability (i.e. the construction of their Lax
operators) is left to an ad hoc procedure. In
\cite{{toppan},{top2},{kst}} it has been
recognized that (bosonic and supersymmetric) Non-Linear
Schr\"{o}dinger-type hierarchies can be obtained from coset
algebra structures. Here it will be shown a method to
systematically produce such kind of hierarchies
and their Lax operators, in terms of
the above mentioned AKS matrix (super)-KP framework
within the homogeneous grading. In particular the supersymmetric case
allows producing generalized super-NLS hierarchies
from any given starting Lie or super-Lie algebra (the Poisson brackets
structure being expressed by the corresponding $N=1$ affinizations).
Now the regular elements belong to the Cartan sector and
are bosonic (while the full Lax operators must be fermionic).
The consistency of this procedure is guaranteed once introduced the
notion of ``twisted" bosonic and fermionic power series in the spectral
parameter $\lambda$ as alternating sums of bosonic-even powers of
$\lambda$ and fermionic odd powers (and conversely); the derived
equations of motion and hamiltonians are of course standard
supersymmetric theories expressed in a manifestly supersymmetric
formalism.
At first sight this construction looks strange, but it is algebraically
perfectly well-defined: it should be also noticed that constructions which
present similar features have already been
encountered in supersymmetric theories, think
for instance to the
GSO projection and to the supersymmetric Witten index \cite{witten}.\par
This paper therefore contains the extension of the
\cite{IK} method to the homogeneous grading case.\par
It should be noticed that, even if a manifestly $N=1$ super-formalism
only has been considered here, $N=2$ supersymmetric hierarchies can be
obtained from the above picture by starting from algebras (and regular
elements) which admit an anti-involution $J$ compatible with the
supersymmetry (see \cite{{hull},{EH}});
this is however a sufficient but not necessary
condition: it has been pointed out in \cite{{ks},{kst}} that already the
standard super-NLS equation, obtained from the $N=1$ affinization of
the $sl(2)$ algebra, admits an $N=2$ structure.
\par
Besides the above construction, the following points will also be
analyzed here: \\
{\em i}) the arising of possible symmetries under
the finite Weyl group or the outer Lie algebra automorphisms
transformations.\\
{\em ii}) The field reductions which can be consistently imposed
once a finite symmetry is present (this is always the case for the
positive versus negative root symmetry).\\
{\em iii}) The iterative prove of the coset structure for the
higher order hamiltonian densities, i.e. their vanishing Poisson
brackets with respect to some affine Lie subalgebra.\\
{\em iv}) The already mentioned role played by different regular elements
to produce different integrable hierarchies.
It is applied in particular to obtain
more general hierarchies,
containing free parameters, than
those studied in \cite{{FK},{AGZ}} (this new situation appears
already from the $sl(3)$ algebra); the
integrable structure is preserved even in presence of these
free parameters.
The generalization with respect to \cite{{FK},{AGZ}}
is due to the fact that the
extra-restriction that the diagonal-transformed Lax operator belongs
to a symmetric space is no longer imposed here (translated into a
geometrical language, this implies not imposing the vanishing of the
torsion).  \\
{\em v}) a heuristic derivation of the
constrained KP-scalar Lax operators from
the matrix ones is also presented.\par
Besides the standard NLS and super-NLS hierarchies obtained from the
affine (and respectively $N=1$ super-affine) $sl(2)$ algebra,
the bosonic
hierarchies derived from the affine $sl(3)$ algebra, as well as
the
supersymmetric hierarchies obtained from the $N=1$ affinization
of the $sl(3)$ algebra and the $osp(1|2)$
superalgebra are also explicitly
presented.\par
A rather surprising result is found: in the $sl(3)$ case,
only the bosonic hierarchies depending on a restrict class of values
for the
free continuous parameter can be supersymmetrically extended
in such a way to lead to an integrable supersymmetric hierarchy.\par
The scheme of this paper is the following:\par
the bosonic AKS approach to matrix-Lax operators is at first recalled.
The coset-structure property of the homogeneous-grading hierarchies
is derived. Then the symmetry properties under Weyl group and outer
automorphisms are analyzed. The $sl(3)$-algebra case will be carried
out completely and the whole set of associated hierarchies
will be written down.
Next, the supersymmetric AKS approach for the homogeneous grading
will be introduced. In particular this scheme will be
applied to explicitly compute the supersymmetric hierarchies associated
to $sl(3)$ and $osp(1|2)$.

\section{Reviewing the AKS framework.}

\indent

In this section I will shortly review the AKS-matrix Lax operator
approach to bosonic integrable hierarchies.
For a more complete account see e.g.
\cite{feh}.\par
As a  starting point a matrix-type Lax operator ${\cal L}$ is assumed,
defined through
\begin{eqnarray}
{\cal L} &=& {{\textstyle{\partial\over \partial x}}} +
J(x) + \Lambda
\label{lax}
\b
Here $x$ is a space coordinate (it can be assumed either $x\in {\bf R}$
or $x\in S^1$). \par
$J(x)$ denotes a set of currents valued in the
semisimple finite Lie algebra ${\cal G}$:
\a
J(x) &=& \sum_i J_i(x) g_i\nonumber\\
\b
where $g_i$'s are the ${\cal G}$-Lie algebra generators
\a
[g_i,g_j] &=& \sum_k{f^k}_{ij} g_k\nonumber\\
\b
and ${f^k}_{ij}$ are the ${\cal G}$-structure constants.\par
The Lie algebra ${\cal G}$ is naturally extended into a loop algebra
${\tilde{\cal G}}$ defined through
\a
{\tilde{\cal G}} &=& {\cal G} \otimes {\bf C} (\lambda, \lambda^{-1})
\b
The elements in ${\tilde{\cal G}}$ are Laurent expansions in the spectral
parameter $\lambda$. The brackets for ${\tilde{\cal G}}$ are given by
\a
[g_i\cdot\lambda^m, g_j\cdot\lambda^n] &=&
\sum_k {f^k}_{ij}g_k\cdot\lambda^{m+n}
\b
for any integers $n,m$.\par
The adjoint operator $ad_Y$ is defined through
\a
ad_Y (X) &=& [Y,X]
\b
where we can assume both $X,Y\in {\cal G}$ or $X,Y\in{\tilde{\cal
G}}$.\par
$\Lambda$ in (\ref{lax}) is a constant (i.e.
not depending on $x$)
regular element of ${\tilde{\cal
G}}$.\par
The regularity has the following meaning: ${\tilde{\cal G}}$
is decomposed as a direct sum
\a
{\tilde{\cal G}} &=& {\tilde{\cal K}} \oplus {\tilde{\cal M}}
\b
where
\a
{\tilde {\cal K}} &=_{def}& Ker(ad_{\Lambda})\nonumber\\
{\tilde {\cal M}} &=_{def}& Im(ad_{\Lambda})
\b
It is furthermore assumed ${\tilde {\cal K}}$ to be abelian;
symbolically
\a
[{\tilde {\cal K}}, {\tilde {\cal K}}] &=&0
\b
In our case the following commutator is satisfied, too:
\a
[{\tilde {\cal K}}, {\tilde {\cal M}}] &\subset& {\tilde {\cal M}}
\b
If, moreover, the following condition is satisfied
\a
[{\tilde{\cal M}},{\tilde {\cal M}}] &\subset & {\tilde {\cal K}}
\label{sym}
\b
then ${\textstyle {{\tilde {\cal G}} \over {\tilde{\cal K}}}}$
is a symmetric space.\par
This case has been studied in \cite{{FK},{AGZ}}, and a full
classification of symmetric spaces is available
(for an account see \cite{per}).\par
In this paper
the most general case, obtained by dropping the condition (\ref{sym}),
will be considered. As a consequence generalizations of the results in
\cite{{FK},{AGZ}} will be obtained.\par
To produce integrable hierarchies the concept of ${\bf Z}$-grading
for the Lie algebra ${\tilde{\cal G}}$ must be introduced. A grading
${\em deg}$ is a linear operator of the form
\a
{\em deg} &=& N\lambda{\textstyle{d\over d\lambda}} + ad_Z
\b
(where $N$ is a non-zero integer and $Z$ is a suitable element in the
Cartan subalgebra)
such that the elements in ${\tilde{\cal G}}$
are eigenvectors of ${\em deg}$ having integer eigenvalues.\par
$\Lambda$ in (\ref{lax}) must be an eigenvector of ${\em deg}$ having
non-zero eigenvalue.\par
I will leave to \cite{feh} the discussion about which are the admissible
gradings for any given Lie algebra. Here I will just remember that any
Lie algebra always admits two extremal gradings (plus, possibly, a series of
intermediate ones): the principal grading and the homogeneous
one. The former associates grade-one to the simple roots of the algebra.
The integrable hierarchies produced from this grading are
generalizations of the KdV equation (the standard
KdV is obtained from the $sl(2)$
algebra, the Boussinesque from $sl(3)$ and so on). \par
The homogeneous grading is defined through
\a
{\em deg} &=& {\textstyle{\lambda }{d\over d \lambda}}
\b
(it counts the
powers in
${\lambda}$). \par
The grade-one regular elements $\Lambda$ have in this case
the form
\a
{\Lambda} &=& \lambda H
\label{regul}
\b
where $H$ is a given, generic element in the Cartan subalgebra of ${\cal
G}$ (such that all its eigenvalues in the adjoint representation of ${\cal
G}$ are different).\par
For this particular grading the decomposition
\a
{\cal G} &=& {\cal K} \oplus {\cal M}
\b
holds, where
\a
{\cal K } &=& Ker (ad_H)\nonumber\\
{\cal M} &=& Im(ad_H)
\b
We have now
\a
{\tilde {\cal K}}&=& {\cal K}\otimes {\bf
C}(\lambda,\lambda^{-1})\nonumber\\
{\tilde {\cal M}} &=& {\cal M} \otimes {\bf C} (\lambda,\lambda^{-1})
\b
The crucial feature of the AKS approach consists in the fact that the
Lax operator ${\cal L}$ provides $(1+1)$-dimensional integrable
hamiltonian systems through the following procedure: at first it should
be noticed that ${\cal L}$
can be diagonalized via a similarity
transformation
\a
{\cal L}\mapsto {\hat {\cal L}}
\b
defined by
\a
{\hat{\cal L}} &=& exp(ad_M)\cdot {\cal L} = \sum_{n=0}^\infty
{\textstyle {1\over n!}}(ad_M)^n ({\cal L})
\label{diag}
\b
where $M$ is a {\it uniquely} defined expansion
of negative-graded elements of ${\tilde{\cal M}}$ which can be
iteratively computed. \par
It turns out that ${\hat{\cal L}}$ is expanded as a sum of
negative-graded diagonal elements of the Lie algebra ${\cal G}$;
they provide an infinite series of mutually commuting
(i.e. having vanishing Poisson brackets), local in the $J_i(x)$ fields,
hamiltonian densities. \par
At least two compatible Poisson brackets
structures can be defined for such systems. Throughout this paper
we will be interested only in the second one, which is given by the
affine-Lie Poisson brackets algebra, defined as the
central extension of the ${\tilde{\cal G}}$-loop algebra.\par
Explicitly we have
\a
\{J_i(x), J_j(y) \} &=& \sum_k {f^k}_{ij} J_k (y) \delta (x-y) + K_{ij}
\partial_y \delta(x-y)
\label{aff}
\b
where in the above formula
\a
K_{ij} &=& Tr( g_ig_j)
\b
in the adjoint representation for ${\cal G}$.\par
In the specific case of the homogeneous grading
the element $M$ in (\ref{diag}) is given by
\a
M &=& \sum_{k=1}^\infty \lambda^{-k} M_k
\b
with
\a
M_k &\in& {\cal M} = Im(ad_H)
\b
The transformed
Lax operator
${\hat {\cal L}}$ is expanded in powers of $\lambda$ as
\a
{\hat{\cal L}} &=& \lambda H +\partial_x +J_\alpha h_\alpha +
\sum_{k=1}^\infty \lambda^{-k} R_{k,\alpha} h_\alpha
\label{laxd}
\b
where $H$ is given by (\ref{regul}) and $h_\alpha$ denote the Cartan
generators of ${\cal G}$
(the sum over $\alpha$ in the above formula is understood).\par
$J_\alpha , R_{k,\alpha}$ for any $k,\alpha$ are mutually commuting
hamiltonian densities which provide the compatible flows associated to
the given hierarchy.

\section{The coset property of the homogeneous hierarchies.}

\indent

In this section it will be proven the coset structure of the hierarchies
associated to the homogeneous grading. More precisely, the following
property is satisfied for the hierarchies determined by a regular
element
\a
\Lambda &=& \lambda\sum_\alpha c_\alpha h_\alpha
\label{cart}
\b
where $h_\alpha$ are the Cartan generators of a given Lie algebra ${\cal
G}$ ($\alpha = 1,2,...,r$, with $r$ the rank of the algebra) and $c_\alpha$
are generic constants.\par
It turns out that with respect to the second Poisson brackets
structure (the affine-Lie Poisson brackets given in (\ref{aff})),
the whole set of higher hamiltonian densities $R_{k,\alpha}$ of
eq. (\ref{laxd}) have vanishing Poisson brackets with respect to the
$J_\alpha(x)$ currents associated to the Cartan generators.\par
Since the $J_\alpha$'s generate independent ${\hat {U(1)}}$
Kac-Moody subalgebras, the above hamiltonian densities
are elements of the ${\hat{\cal G}}$-enveloping algebra which belong
to the \\
${\hat {U(1)}}^r= {\hat{U(1)}}\otimes ...\otimes {\hat{U(1)}}$
 ($r$ times) coset subsector.\par
The above property is satisfied for generic $c_\alpha$ in (\ref{cart}).
For some specific values of $c_\alpha$ the coset algebra can be bigger
and coincide with some non-abelian Kac-Moody subalgebra of ${\hat {\cal
G}}$.
\par
In order to prove the above theorem it is convenient to introduce
in full generality \cite{{top1},{top2},{toppan}} the notion
of charged fields and covariant derivative with respect to the
${\hat{U(1)}}$ Kac-Moody algebra defined by the brackets
\a
\{J_0(x),J_0(y)\}&=&{\textstyle{\partial\over\partial y}}\delta(x-y)
\b
A $q$-charged field $V_q$ is defined to satisfy
\a
\{J_0(x),V_q(y)\}&=& qV_q(y)\delta(x-y)
\b
while a covariant derivative ${\cal D}$ acting on $V_q$ can be
introduced through the position
\a
{\cal D} V_q (x) &=_{def}& (\partial + q J_0(x)) V_q(x)
\b
Notice that the covariant derivative maps $q$-charged fields into new
fields of definite charge having the same value $q$.\par
The coset property of the $R_{k,\alpha}$ hamiltonian densities implies
that they are chargeless differential polynomials constructed with
the subset of $J_i(x)$'s given by charged fields
and covariant derivatives acting on them.\par
Such a theorem can be easily proven by using an iterative procedure.
For simplicity it will be given for the $sl(2)$ case (which allows
also to introduce the standard Non-Linear-Schr\"{o}dinger equation), the
generalization to generic Lie algebras ${\cal G}$ is
straightforward.\par
The $sl(2)$ algebra is generated by $H, E_\pm$, satisfying the
commutation relations:
\a
\relax{[H, E_\pm ]}&=& \pm 2 E_\pm \nonumber\\
\relax{[E_+,E_-]}&=& H
\label{sl2algebra}
\b
The associated second (affine-Lie) Poisson brackets structure is
expressed by
\a
\{J_0(x),J_0 (y)\}&=& \partial_y\delta(x-y)\nonumber\\
\{J_0(x),J_\pm (y)\}&=& \pm 2 J_\pm(y)\delta(x-y)\nonumber\\
\{J_+(x), J_-(y)\}&=& 2{\cal D}_y\delta(x-y) = 2(\partial_y\delta(x-y)-
2J_0(y)\delta(x-y))
\label{sl2aff}
\b
Here $J_\pm(x)$ have charge $\pm 2$ with respect to the ${\hat{U(1)}}$
subalgebra generated by $J_0(x)$.
\par
Any other Poisson bracket is vanishing.
\par
The Lax operator ${\cal L}$ is given by
\a
{\cal L} &=& \partial + J_o(x) H + J_+(x) E_+ + J_-(x)E_- + \lambda H
\label{matr}
\b
We can diagonalize, order by order in $\lambda$,
the above Lax operator into ${\hat{\cal L}}$ such that
\a
{\hat{\cal L}} &=& exp(ad_M) ({\cal L}) = \partial + J_0(x) H +
\sum_{k=1}^\infty \lambda^{-k} R_k H
\b
where the diagonalizing matrix has the form
\a
M &=& \sum_{i=1}^\infty \lambda^{-i}(M_{i,+} E_+ + M_{i,-}E_- )
\label{sl2dia}
\b
At the lowest orders we find
\a
M_{1\pm} &=& \pm J_\pm\nonumber\\
M_{2,\pm} &=& -{\cal D} J_\pm \nonumber\\
M_{3,\pm} &=& \pm ({\cal D}^2 J_\pm -{\textstyle{4\over 3}}(J_+J_-) J_\pm
\b
while the hamiltonian densities are given by
\a
R_1 &=& J_+J_-\nonumber\\
R_2 &=&{\textstyle{1\over 2}} (J_+{\cal D} J_- -J_-{\cal D} J_+ )
\b
At the lowest orders $M_{i,\pm}$ are differential polynomials with
definite charge $\pm 2$ respectively, while the hamiltonian densities
are chargeless differential polynomials.\par
It is immediately shown, due to the properties of the adjoint action
$exp(ad_M)$ acting on ${\cal L}$, that assuming $M_{i,\pm}$ having
charge $\pm 2$ for $i=1,...,N$ and $R_i$ being chargeless for
$i=1,...,N$, necessarily follows that $M_{N+1,\pm}$ have charges $\pm 2$
and $R_{N+1}$ is chargeless. The theorem is therefore proven by
induction. Its generalization to arbitrary Lie algebras is
straightforward.
\par
The different hamiltonian flows for integral values $k=1,2,... $
are defined through the following
equation, for any given field $\phi(x)$:
\a
{\textstyle{\partial\over \partial t_k}} \phi (x) &=& {\textstyle
{1\over 2}} \{ \phi (x), \int dy R_k (y) \}
\b
(the factor $2$ is introduced for normalization convenience).\par
Since $J_0(x)$ has vanishing Poisson brackets with respect to any $R_k$,
we get for any flow
\a
{\textstyle {\partial\over\partial t_k}}J_0(x) &=& 0
\b
It follows that it is consistent with the equations of motion to set
\a
J_0&\equiv &0
\label{jzero}
\b
In literature the above position is in general set as a Dirac constraint.
In our approach it is recovered as a consequence of the equations of motion,
which implies a simplified analysis (in particular it avoids computing
Dirac's brackets to obtain the flows, which is of great
help
in many cases, see e.g. \cite{kst}).
\par
The first two flows for the fields $J_\pm(x)$ are respectively
given by
\a
{\textstyle{\partial\over\partial t_1}} J_\pm (x) &=& -
{\cal D} J_\pm
(x)
\b
and
\a
{\textstyle{\partial\over \partial t_2}}
J_\pm (x) &=& \pm ( {\cal D}^2 J_\pm (x)
+ 2(J_+J_-)J_\pm (x))
\b
The second flow is precisely the two-components
Non-Linear-Schr\"{o}dinger equation. \par
The covariant derivative
in the above formulas can be replaced by the standard derivative,
once setting the (\ref{jzero}) solution to the equations of motion.

\section{From matrix to scalar Lax operators: a heuristic derivation.}

\indent

Before going ahead, let me just point out that a connection exists
between matrix-type Lax operators and consistent field-restrictions
of the scalar KP operator. A detailed analysis has been given in
\cite{AGZ}. Here I wish just furnish a simple heuristic argument
to understand such a connection. For simplicity I will treat the
$sl(2)$ case in the homogeneous grading; the extension to generic
algebras
can also be given along the same lines.\par
Let us consider as a starting point the equation
\begin{eqnarray}
{\cal L} \cdot\Psi &=& 0
\end{eqnarray}
(where ${\cal L}$ is the matrix Lax operator (\ref{matr}))
in some given representation of the $sl(2)$ algebra.\par
The $\lambda\equiv 0$ component of the above equation in the fundamental
(spin${\textstyle{1\over 2}}$) representation for $sl(2)$ gives us:
\begin{eqnarray}
\left(\partial +
\left( \begin{array}{cc}
J_0 & J_+ \\
J_- & -J_0 \end{array} \right)\right)
\left( \begin{array}{c} \Psi_+\\ \Psi_- \end{array}\right) &=&0
\end{eqnarray}
If we solve the above equation for, let's say, the $\Psi_-$ component and
allow formally inverting the derivative operator, then we can plug the result
into the equation for the $\Psi_+$ component, obtaining:
\begin{eqnarray}
({\cal D} + J_- {\cal D}^{-1} J_+ ) \Psi_+ = 0
\quad&\equiv&\quad L\cdot \Psi_+=0
\end{eqnarray}
The scalar operator
\begin{eqnarray}
L&=& {\cal D } +J_-{\cal D}^{-1}J_+
\end{eqnarray}
(It also turns out $L\equiv \partial + J_-\partial^{-1}J_+$ when inserting
the constraint, compatible with the equations of motion, $J_0 =0$)
provides the consistent field reduction of the scalar KP operator
associated to the $2$-component NLS equation (see \cite{top2}).\par
It should be noticed that to a given matrix-type Lax operator
one can associate different but equivalent scalar KP restrictions,
according to which representation of the algebra has been chosen.
For instance, if instead of starting with
the fundamental representation
of $sl(2)$ we proceed from the triplet representation (acting on the
vector $(\Psi_1, \Psi_0,\Psi_{-1}$)) we are led,
after solving the equations for
the $\Psi_{\pm 1}$ components, to the following relation:
\begin{eqnarray}
({\cal D} +J_-{\cal D}^{-1}J_+
+J_+{\cal D}^{-1}J_- ) \Psi_0 &=& 0
\label{scalar}
\end{eqnarray}
The new scalar Lax operator $L'$
\begin{eqnarray}
L' &=&
{\cal D} +J_-{\cal D}^{-1}J_+
+J_+{\cal D}^{-1}J_-
\end{eqnarray}
is equivalent to the $L$ Lax operator (\ref{scalar}) since
their hamiltonian densities differ by total derivatives.
$L'$ is basically the symmetrized form of $L$ under the exchange
$J_-\leftrightarrow J_+$.

\section{Field reductions: the positive-negative roots exchange.}

\indent

In the following sections the symmetries under Weyl group
transformations and outer automorphisms for generic Lie algebras will be
analyzed. Here I will study the simplest such kind of symmetries,
already appearing for the $sl(2)$ algebra: the algebra
automorphism which exchanges
positive and negative roots. Such ${\bf Z}_2$
symmetry is always present for any
Lie algebra; in general it will be provided by a combination of
a Weyl transformation and an outer automorphism; for the $sl(2)$
algebra, which does not admit outer automorphisms, it coincides with the
(unique) Weyl transformation. \par
Explicitly we have
\a
E_+&\leftrightarrow  &E_-; \quad\quad H\mapsto -H
\b
Such transformation can be extended to the affine-Lie automorphism
\a
J_+(x) &\leftrightarrow& J_-(x);\quad\quad J_0(x)\mapsto -J_0(x)
\b
where $J_0(x), J_\pm (x)$ generates the  ${\hat{sl(2)}}$ algebra given
in
(\ref{sl2aff}).\par
The   ${\cal L}=\partial + J_0(x) H + J_+(x)E_+ +J_-(x) E_- + \lambda H $
\\
Lax operator of (\ref{matr}) is invariant under the above
transformation,
provided that the spectral parameter $\lambda $ being transformed
according to
\a
\lambda &\mapsto& -\lambda
\label{l-l}
\b
Moreover, it is easily realized that under the above
transformation
the diagonalizing matrix $M$ in (\ref{sl2dia}) is left
invariant. As a consequence
the diagonalized Lax operator ${\hat{\cal L}}$ itself is invariant.\par
Due to the $\lambda$-transformation property (\ref{l-l}) the
odd hamiltonian densities $R_k$ ($k$ odd)
are left invariant, while the even ones ($R_k$ with $k$ even) are
transformed into their opposite:
\a
R_k\mapsto (-1)^{k+1} R_k
\b
Each time we have a symmetry we can perform a field reduction,
identifying the fields which are related by the symmmetry
transformation. Such a step can be performed for our ${\bf
Z}_2$-symmetry, which allows us passing from the $2$-component NLS equation
to the standard-form single component Non-Linear-Schr\"{o}dinger
equation. However it is worth to notice the following point:
the hamiltonian
which generates the NLS equation is the second one ($R_2$) which is not
invariant under the symmetry, but it is transformed into its opposite.
For that reason it is not possible to identify $J_+$ with $J_-$, instead
we have to assume the time $t_2$ being imaginary ($t_2=it$) and
\a
J_+(x) &=& u(x) = {J_-}^\star(x)
\b
The situation here is parallel to what happens in quantum-mechanics
when disposing of a time reversal
transformation which flips the sign of the hamiltonian:
in that case the symmetry is recovered in terms of an antiunitary
transformation which involves complex conjugation.\par
The final result for the single-component NLS equation is the following:
\a
i{\dot u} &=& u'' + 2u|u|^2
\b
(here the standard convention of denoting time and spatial derivatives
with respectively a dot or a prime has been adopted).

\section{The dependence on the regular element and its symmetry
properties.}

\indent

In this section I wish to analyze a new feature, not touched by
our previous discussion, that is the dependence of the integrable
hierarchies from the choice of the regular element and its symmetry
properties. For the $sl(2)$ algebra case this problem can not be posed
since the grade-one regular element with respect to the homogeneous
grading (that is $\lambda H$ of (\ref{regul}))
is essentially unique (apart an overall normalization factor
which can be reabsorbed by rescaling the spectral parameter).\par
For general Lie algebras the problem of determining which different
hierarchies are produced from different regular elements is a very
interesting one. To be definite here we analyze
in full detail the $sl(3)$ algebra
case. This is indeed a very fundamental case because it already contains
all the features (namely a non-trivial Weyl group and the presence
of an outer automorphism) which are found in more complicated
examples for generic Lie algebras.
The generalization of the approach here developed
to such cases is immediate, it is only technically more involved.\par
Before introducing my conventions concerning the $sl(3)$ algebra
let me just recall (see \cite{gil} for a complete account) that the Weyl
group associated to a given Lie algebra is a finite group of reflections
which leave invariant the root systems of the algebra. It coincides with
a subgroup of the inner automorphisms of the Lie algebra.\par
Besides the
inner automorphisms a generic Lie algebra admits also a group of outer
automorphisms (i.e. they can not be obtained as an Adjoint action\par
$x \mapsto x' = exp(ad_y)(x)$, with $x,y,x'\in {\cal G}$)\\
which coincides with the group of symmetries of its Dynkin diagram.\par
The $sl(3)$ algebra admits $8$ generators. $2$ generators, denoted
as $H_1$ and $H_2$, belong to the Cartan sector (rank $2$); the simple
(positive and negative) roots will be denoted as $E_{\pm1}$, $E_{\pm2}$
respectively. The extra (maximal) root will be represented as
$E_{\pm3}$.\par
It is convenient to introduce the $3\times 3$ matrices $e_{ij}$, for
$i,j=1,2,3$, defined as follows:
$e_{ij} $ has all zero entries apart $1$ in the $i$-th raw, $j$-th
column position.\par
The fundamental $3\times 3$ representation of $sl(3)$ is obtained
by setting
\a
H_1&=& e_{11} - e_{22},\quad\quad\quad H_2= e_{22}-e_{33},
\b
for the Cartan generators,
\a
E_{+1} &=& e_{12},\quad\quad E_{+2}=e_{23},\quad\quad E_{+3}=e_{13},
\b
for the positive roots and
\a
E_{-1}&=& e_{21},\quad\quad E_{-2}=e_{32},\quad\quad E_{-3}=e_{31},
\b
for the negative ones.\par
The full commutation relations of the $sl(3)$ algebra can be easily
computed from the above positions.\par
The Weyl group for $sl(3)$ coincides with the $S_3$
permutation group (of
order $6$) of three elements denoted as $e_1,e_2,e_3$. \par
The positive roots can be associated to the following combinations of
$e_i$'s (see \cite{gil}):
\a
E_{+1}&\equiv & e_1-e_2; \quad\quad E_{+2} \equiv e_2-e_3;\quad\quad
E_{+3}\equiv e_1-e_3
\b
The Weyl group admits $3$ distinct ${\bf Z}_2$ subsymmetries
$s_i$, $i=1,2,3$, given by the corresponding reflections
along the $e_i$ element in $S_3$,
acting as:
\a
s_1&:& E_{\pm 1}\leftrightarrow E_{\pm 3}; \quad\quad
E_{+2}\leftrightarrow E_{-2};\quad\quad H_1\mapsto H_1 + H_2; \quad\quad
H_2\mapsto -H_2.\nonumber\\
s_2&:& E_{\pm 1} \leftrightarrow E_{\mp 2};\quad\quad E_{+3}
\leftrightarrow E_{-3};\quad\quad H_1\leftrightarrow -H_2.\nonumber\\
s_3&:& E_{+1}\leftrightarrow E_{-1};\quad\quad E_{\pm2}\leftrightarrow
E_{\pm3};\quad\quad H_1\mapsto -H_1;\quad\quad H_2\mapsto H_1+H_2.
\b
The ${\bf Z}_3$ subsymmetry obtained by sending $1\mapsto 2\mapsto
3\mapsto 1$ leads to
\a
&& E_{\pm1}\mapsto E_{\pm 2}\mapsto E_{\mp 3}\mapsto E_{\pm
1};\nonumber\\
&& H_1\mapsto H_2\mapsto -(H_1+H_2).
\b
Besides the above Weyl transformation, an extra ${\bf Z}_2$ symmetry
is present: it is realized by the outer automorphism $\sigma$ which exchanges
the two simple roots.
It is explicitly given by the following relations
\a
{\sigma}&:& E_{\pm 1}\leftrightarrow E_{\pm2};\quad\quad E_{\pm
3}\mapsto - E_{\pm 3}; \quad\quad H_1\leftrightarrow H_2.
\label{sigma}
\b
It should be noticed that the automorphism $s_{\pm}$ which exchanges
positive and negative roots is in this case given by the combination of
the $s_2$ Weyl transformation and the outer automorphism $\sigma$:
\a
s_\pm &=& s_2\cdot \sigma
\b
Explicitly we have
\a
s_\pm&:& E_{+1}\leftrightarrow E_{-1};\quad\quad E_{+2}\leftrightarrow
E_{-2};\quad\quad E_{+3}\leftrightarrow - E_{-3};\quad\quad
H_1\leftrightarrow -H_1;\quad\quad H_2\leftrightarrow -H_2.\nonumber\\
&&
\b
Both the Weyl transformations and the $\sigma$ outer automorphism can be
extended to be automorphisms for the full affine ${\hat{sl(3)}}$
algebra, in precise analogy to what discussed in the previous
section.\par
The generic grade-one $sl(3)$ regular element
$\Lambda$ for the homogeneous grading has the following form
\a
\Lambda &=&\lambda H = \lambda ( {cos^2\theta} H_1 +{sin^2{\theta}} H_2
)
\label{lambdaang}
\b
Therefore it will depend on an arbitrary angle $\theta$ (as in the
$sl(2)$ case an overall normalization factor can be reabsorbed in the
definition of $\lambda$). \par
It will be explained in the next section that the integrable hierarchies
have an essential dependence on $\theta$, that is $\theta$ can not be
rescaled at will.\par
As already recalled, the AKS framework works if the eigenvalues
of the regular element are all distinct. In the $3\times 3$ fundamental
representation the diagonal for $H$ in (\ref{lambdaang}) is
given by
\a
&&( t, 1-2t, t-1)
\b
(where for simplicity we have set $t={cos^2\theta}$, $t\in [0,1]$).\par
It follows that two values exist
\a
&& t={\textstyle{1\over 3}};\quad\quad t={\textstyle{2\over 3}}
\label{special}
\b
which must be excluded. \par
However it will be shown later that both the
(\ref{special}) conditions
still produce admissible integrable hierarchies (of degenerate type);
they have been discussed in \cite{AGZ}.\par
Let us discuss now the symmetry properties under Weyl transformations
and outer automorphism $\sigma$ for the matrix Lax operator ${\cal L}$
(and therefore of its associated hierarchies)
according to the choice of the regular element $\Lambda$. \par
We have for $sl(3)$
\a
{\cal L} &=& \partial + \sum_i J_i(x) g_i + \lambda( t H_1 + (
1-t) H_2)
\label{lax3}
\b
The term $\sum_i J_i(x) g_i $ (the sum is over the $sl(3)$ generators)
is invariant under both the Weyl and the $\sigma $ transformations due to the
combined transformation properties of $g_i, J_i(x)$; obviously
the derivative $\partial$
is invariant too.\par
For what concerns $\Lambda$, the transformations act as follows:\\
{\em i}) $\sigma$ maps ${\cal L} (t)$ into ${\cal L} (1-t)$. As a consequence
$t$ and $1-t$ produce the same set of hamiltonian densities
and therefore the same hierarchies. There is only one symmetric point
\a
t&=& {\textstyle{1\over 2}}
\b
which leaves ${\cal L}$ invariant.\par
It corresponds to the choice
\a
\Lambda &\equiv& \lambda ( {\textstyle{1\over 2}},0,-{\textstyle{1\over
2}})
\b
on the diagonal. This value of $t$ allows the folding procedure
(see \cite{fold}) which will be discussed in more detail in
the next section.\\
{\em ii}) the $s_2$ transformation is a symmetry for ${\cal L}$ only for
$t={\textstyle{1\over 2}}$ and assuming $\lambda$ to be mapped into
its opposite ($\lambda\mapsto-\lambda$).\\
{\em iii}) The combined $s_\pm = s_2\cdot \sigma$ transformation
(positive-negative roots exchange) is a symmetry of ${\cal L}$ for any
value of $t$, provided that $\lambda\mapsto -\lambda$
under $s_\pm$. As a consequence, for any value of $t$ (or of the
$\theta$-angle), the reduction from the $2$-component fields hierarchy
to the single(complex)-component fields hierarchy can be performed. The
same remarks concerning the alternate parity of the
hamiltonian in the $sl(2)$ case hold here as well.\\
{\em iv}) For what concerns the $s_1$ transformation, it can act as
a symmetry for
${\cal L}$ if one of the two conditions below is satisfied:\\
{\em either} $\lambda$ is unchanged ($\lambda\mapsto \lambda$) under
$s_1$;
in this case $t$ must assume the degenerate value $t={\textstyle {2\over 3}}$
so that
\a
\Lambda &\equiv& \lambda ( {\textstyle {2\over 3}}, -{\textstyle{1\over
3}},-{\textstyle{1\over 3}})
\b
on the diagonal,\\
{\em or} $\lambda$ is mapped into its opposite
($\lambda\mapsto-\lambda$) and $t=0$; therefore
\a
\Lambda &\equiv& \lambda (0,1,-1)
\b
on the diagonal.\\
{\em v}) The case concerning the $s_3$ symmetry is specular to the
previous one, to which it can be reduced after performing a  $\sigma$
transformation. ${\cal L}$ is symmetric under $s_3$ if either\par
$s_3: \lambda\mapsto\lambda$ and $t={\textstyle{1\over 3}}$, or \par
$s_3: \lambda\mapsto -\lambda $ and $t=1$.\par
It is not necessary to analyze the transformation properties for other
elements of the Weyl group since the latter, being a permutation group,
is recovered from the application of two generators (which can be
assumed to be e.g. $s_1,s_2$).

\section{The $sl(3)$ hierarchies in the homogeneous grading.}

\indent

In this section the results previously obtained will be applied to
construct the whole set of integrable hierarchies associated to the
$sl(3)$ algebra in the homogeneous grading.\par
At first it is convenient to explicitly introduce the affine ${\hat
{sl(3)}}$
algebra
which provides the second Poisson brackets structure.\par
The two currents $J_{0,1}(x), J_{0,2}(x)$ are associated with the two
generators in the Cartan subalgebra, while the positive (negative)
roots correspond to the currents $J_{\pm i}$, $i=1,2,3$.\par
The Cartan subalgebra reads as follows
\a
\{J_{0,1}(x),J_{0,1}(y)\} &=& 2\partial_y\delta(x-y)\nonumber\\
\{J_{0,1}(x),J_{0,2}(y)\} &=& -\partial_y\delta(x-y)\nonumber\\
\{J_{0,2}(x),J_{0,2}(y)\} &=& 2\partial_y\delta(x-y)
\label{aff30}
\b
The currents $J_{\pm i}(x)$ are charged fields
\a
\{J_{0,j}(x), J_{\pm i}(y)\}&=& q_{\pm i,j} J_{\pm i}(y)\delta(x-y)
\label{aff31}
\b
with charges $q_{\pm i}\equiv (q_{\pm i, 1}, q_{\pm i, 2})$ given by
\a
q_{\pm 1} &=& \pm (2,-1)\nonumber\\
q_{\pm 2} &=& \pm (-1,2)\nonumber\\
q_{\pm 3} &=& \pm (1,1)
\b
The covariant derivatives turn out to be
\a
{\cal D} J_{\pm 1} &=& \partial J_{\pm 1} \mp J_{0,1} J_{\pm
1}\nonumber\\
{\cal D} J_{\pm 2} &=& \partial J_{\pm 2} \mp J_{0,2} J_{\pm
2}\nonumber\\
{\cal D} J_{\pm 3} &=& \partial J_{\pm 3} \mp (J_{0,1} + J_{0,2}) J_{\pm
3}
\b
The algebra is completed by the following relations
\a
\{J_{+i}(x), J_{-i}(y)\} &=& {\cal D}_y \delta (x-y)\quad\quad\quad
for\quad i=1,2,3.\nonumber\\
\{J_{+1} (x), J_{-3}(y)\} &=&
-J_{-2}(y)\delta(x-y)\quad\quad\quad
\{J_{+2}(x),J_{-3}(y)\} = J_{-1}(y) \delta(x-y)\nonumber\\
\{J_{+3}(x),J_{-1}(y)\}&=&-J_{+2}(y)\delta(x-y)\quad\quad\quad
\{J_{+3}(x),J_{-2}(y)\} = J_{+1}(y)\delta(x-y)\nonumber\\
\{J_{\pm 1}(x), J_{\pm 2}(y)\}&=& \pm J_{\pm 3}(y)\delta (x-y)
\label{aff32}
\b
Any other Poisson bracket is vanishing.\par
We have now all the ingredients to compute the integrable
hierarchies.\par
The Lax operator ${\cal L}$ is given in (\ref{lax3}) and depends on
the parameter $t={cos^2\theta}$. It is diagonalized with a similarity
transformation into ${\hat{\cal L}}$:
\a
{\hat{\cal L}} &=& exp(ad_M)({\cal L} =
\nonumber\\
&=&\lambda( tH_1 +(1-t)H_2) +\partial + J_{0,1}(x)H_1+J_{0,2}(x)H_2
+\sum_{k,\alpha}\lambda^{-k} R_{k,\alpha}(x) H_{\alpha}
\b
where $k=1,2,...$ denotes positive integers and $\alpha=1,2$.
\par
The diagonalizing matrix $M$ can be expanded
as
\a
M&=&\sum_{i=1}^\infty(\lambda^{-i}\cdot \sum_j M_{i, \pm j}
E_{\pm j})\b
for $j=1,2,3$.\par
The compatible hamiltonian densities $R_{k,\alpha}$ can be computed with
straightforward techniques. We get at the lowest orders:
\a
R_{1,1} (x)&=& (J_{+3}J_{-3} +{\textstyle{1\over
(3t-1)}}J_{+1}J_{-1})(x)\nonumber\\
R_{1,2} (x) &=& (J_{+3} J_{-3} -
{\textstyle{1\over(3t-2)}}J_{+2}J_{-2})(x)
\label{hami1}
\b
and
\a
R_{2,1} (x) &=&
{\textstyle{1\over 2}}
({\cal D}J_{+3}J_{-3} -J_{+3}{\cal D}J_{-3})  + {\textstyle {1\over
2(3t-1)^2}}({\cal D}J_{+1}J_{-1}
-J_{-1}{\cal D}J_{+1})\nonumber\\
&& +{\textstyle {1\over (3t-1)}} (
J_{+3}J_{-1}J_{-2} +J_{-3}J_{+1}J_{+2})\nonumber\\
R_{2,2} (x) &=& {\textstyle{1\over 2}}
({\cal D }J_{+3}J_{-3}-J_{+3}{\cal D}J_{-3}) +
{\textstyle{1\over 2(3t-2)^2}}({\cal
D}J_{+2}J_{-2}-J_{+2}{\cal D}J_{-2})\nonumber\\
&& +{\textstyle{1\over
(3t-2)}}(J_{+3}J_{-1}J_{-2}+J_{-3}J_{+1}J_{+2})
\label{hami2}
\b
Notice that the hamiltonian densities $R_{k,2}$ are just obtained by
applying a
$\sigma$ transformation (\ref{sigma}) to $R_{k,1}$ (and conversely);
here
\a
\sigma&:& J_{\pm 1}\leftrightarrow J_{\pm 2};\quad\quad
J_{\pm 3} \mapsto -J_{\pm 3}; \quad\quad t\mapsto (1-t).
\b
The hamiltonians $H_{k,\alpha}$ are defined as the integrals
\a
H_{k,\alpha} &=& \int dy R_{k,\alpha}(y)
\b
and the corresponding flows, for a generic field $\phi (x)$, are given
by
\a
{\partial\over\partial t_{k,\alpha}} \phi (x) &=& \{ \phi (x),
H_{k,\alpha}\}
\b
where in the right hand side we have the affine Lie Poisson brackets
(\ref{aff30},\ref{aff31},\ref{aff32}).\par
In the above formulas (\ref{hami1},\ref{hami2})
$t$ can assume any value $t\in [0,1]$ apart
$t={\textstyle {1\over 3}},{\textstyle{2\over 3}}$.\par
Neverthless even
in such degenerate cases we obtain integrable hierarchies. Indeed
for $t={\textstyle{1\over 3}}$ the integrals $H_{k,1}$
are not defined. However, the subset of integrals
$H_{k,2}$, $k=1,2,...$, is well-defined; they provide the
mutually commuting hamiltonians with respect to the second Poisson
brackets structure (\ref{aff30},\ref{aff31},\ref{aff32}).\par
It can be easily shown that in this case it is consistent with the
whole set of $t_{k,2}$ flows not only to put the constraint
\a
J_{0,1}(x)&=&J_{0,2}(x)=0
\b
(see the discussion in section $3$), but also to set
\a
J_{\pm 1} (x) &=& 0
\b
(the consistency of this position is due to the fact that, in the
equations of motion for $J_{\pm 1}(x)$, the right hand side is
proportional to $J_{\pm 1}$).\par
The hierarchy so derived will depend on the fields $J_{\pm 2}(x), J_{\pm
3}(x)$ only.\par
The hamiltonians $H_{k,2}$ belong to the symmetric coset
space ${\textstyle{sl(3)\over sl(2)\times U(1)}}$
(it should be noticed however that the hamiltonian densities
$R_{k,2}$ do not belong to the full affine coset subspace
${{{\hat {sl(2)}}\times {\hat{U(1)}}}}$).\par
Similar considerations hold for $t={\textstyle{2\over 3}}$, but now we
have to
replace $1\leftrightarrow 2$ in the discussion above.\par
Another value of $t$ which must be singled out is $t={\textstyle{1\over
2}}$; it corresponds to the symmetric point which leaves
${\cal L}$ invariant under the outer automorphism $\sigma$. \par
In this case it is convenient to reexpress the fields in terms of the
$\sigma$-eigenvectors: we have as eigenvectors corresponding
to the $(+1)$ eigenvalue:
\a
J_{\pm up} &=& J_{\pm 1} + J_{\pm 2}
\b
while the $(-1)$ eigenvectors
are
\a
J_{\pm down} &=& J_{\pm 1} - J_{\pm 2}; \quad\quad J_{\pm 3}.
\b
A consistent reduction of the ($t={\textstyle{1\over
2}}$) symmetric hierarchy can be obtained by considering the
subset (for positive integers $k$) of
\a
H_{k, up} &=_{def}& H_{k,1} + H_{k,2}
\b
$(+1)$ eigenvectors
hamiltonians and setting all the fields corresponding to the $(-1)$
eigenvalue equal to zero:
\a
J_{\pm down} &=& J_{\pm 3} =0
\b
The reduced hierarchy is just the NLS-hierarchy of section 3.\par
The above procedure is nothing else than a simple example of folding,
that is the reduction associated to a symmetry of the Lie algebra Dynkin
diagram. For general folding constructions see \cite{fold}.\par
Let us come back now to the general case (corresponding to
generic values of $t$).\par
The hierarchies will depend on the whole set of fields $J_{\pm i}$,
$i=1,2,3$ while, as before, we can consistently set
\a
J_{0,1}&=&J_{0,2}=0
\b
at the level of the equations of motion (i.e.
after computing the Poisson brackets).\par
{}From the hamiltonian $H_{1,1}$ we obtain the flow
\a
{\dot J_{\pm 1}} &=& \mp J_{\pm 3} J_{\mp 2} - {\textstyle{1\over
(3t-1)}} {J_{\pm 1}}'\nonumber\\
{\dot J_{\pm 2}} &=& \pm
(1-{\textstyle{1\over (3t-1)}})J_{\pm 3}J_{\mp
1}\nonumber\\
{\dot J_{\pm 3}} &=& \mp {\textstyle {1\over (3t-1)}} J_{\pm1}J_{\pm 2}
- {J_{\pm 3}}'
\label{firstflow}
\b
The flow associated to $H_{1,2}$ is obtained from (\ref{firstflow}) by
replacing $t\mapsto (1-t)$ and $1\leftrightarrow 2$
(for any couple of $k$-th order hamiltonians
$H_{k,1}$, $H_{k,2}$ such a replacement obviously holds).\par
{}From the second order hamiltonian $H_{2,1}$ we get the flow
\a
{\dot J_{\pm 1}}&=&
-\gamma J_{\pm 3}J_{\mp 2}' -(1+\gamma)J_{\pm 3}'J_{\mp 2}
\mp \gamma^2 J_{\pm 1}''+\nonumber\\
&& \pm J_{\pm 1} [2\gamma^2 J_{+1}J_{-1}-\gamma J_{+2}J_{-2} +(1+\gamma)
J_{+3}J_{-3}]\nonumber\\
&&\nonumber\\
{\dot J_{\pm 2}}&=&\gamma(\gamma-1)J_{\pm 3}J_{\mp 1}' +(1-\gamma)
J_{\pm 3}' J_{\mp 1}+
\nonumber\\
&& \pm J_{\pm 2} [\gamma(1-\gamma) J_{+1}J_{-1}
+(1-\gamma)J_{+3}J_{-3}]\nonumber\\
&&\nonumber\\
{\dot J_{\pm 3}}&=&\mp J_{\pm 3}'' -\gamma(\gamma +1)J_{\pm 1}'J_{\pm 2}
-\gamma J_{\pm 1}J_{\pm 2}' +\nonumber\\
&& \pm J_{\pm 3} [ \gamma(\gamma+1) J_{+1}J_{-1}-\gamma J_{+2}J_{-2}
+2J_{+3}J_{-3}]
\label{secondflow}
\b
where, in order to simplify our notation, we have set
\a
\gamma &=_{def} & {1\over (3t-1)}
\b
The above (\ref{secondflow}) relations provide the $sl(3)$ generalization of
the
$2$-component fields NLS equation.\par
Due to the results of the previous section concerning the $\pm$ roots
exchange symmetry $s_\pm$, the above relations can be
consistently reduced to the
equations of motion for single-component complex fields;
for the first flow the identification implies
\a
\phi_1(x) &=& J_{+1}=J_{-1}\nonumber\\
\phi_2(x) &=& J_{+2}=J_{-2}\nonumber\\
\phi_3(x) &=& J_{+3} = -J_{-3}
\b
We obtain as a consequence
\a
{\dot{\phi}}_1 &=& -\phi_2\phi_3-\gamma \phi_1'\nonumber\\
{\dot{\phi}}_2 &=& (1-\gamma) \phi_1\phi_3\nonumber\\
{\dot{\phi}}_3 &=& -\gamma\phi_1\phi_2 - \phi_3'
\label{firstflowsusy3}
\b
To get the second flow restriction we must
let the time being
imaginary and set:
\a
\phi_1(x) &=& J_{+1} = J_{-1}^\star;\nonumber\\
\phi_2(x) &=& J_{+2} = J_{-2}^\star;\nonumber\\
\phi_3(x) &=& J_{+3} = - J_{-3}^\star
\b
The second flow provides the generalization of the single-component
NLS equation:
\a
i{\dot\phi}_1 &=& -\gamma\phi_3{{\phi_2}^\star}'-(1+\gamma)\phi_3'
{\phi_2}^\star -\gamma^2 \phi_1'' +\nonumber\\
&& +\phi_1 [ 2\gamma^2
|\phi_1|^2-\gamma|\phi_2|^2-(1+\gamma)|\phi_3|^2]\nonumber\\
i{\dot\phi}_2 &=& \gamma(\gamma-1) \phi_3{{\phi_1}^\star}'
+(1-\gamma)\phi_3'{\phi_1}^\star +\nonumber\\
&& +\phi_2[\gamma(1-\gamma)|\phi_1|^2+(\gamma-1)|\phi_3|^2]\nonumber\\
i{\dot\phi}_3 &=& -\phi_3''-\gamma(\gamma +1)\phi_1'\phi_2
-\gamma\phi_1\phi_2' +\nonumber\\
&&+\phi_3[\gamma(\gamma+1)|\phi_1|^2-\gamma|\phi_2|^2-2|\phi_3|^2]
\label{sl3unicomp}
\b
Since the $s_\pm$ symmetry commutes with the $s_1$, $s_3$
symmetries (see the previous section), the reduction from $2$-component
fields to single-component fields can be performed also in the case
of ``degenerate" hierarchies for $t={\textstyle{1\over
3}}$,${\textstyle{2\over 3}}$.\par
Let us make now some comments about the (\ref{secondflow}) hierarchy. The
parameter $\gamma$ can assume
the real values
\a
\gamma &\geq&{\textstyle {1\over 2}}\quad\quad for\quad {\textstyle
{1\over 3}}< t\leq 1
\label{gamone}
\b
and
\a
\gamma &\leq& -1\quad\quad for\quad 0\leq t<{\textstyle{1\over 3}}
\label{gamtwo}
\b
Under the $\sigma$ transformation $\gamma$ is mapped into
${\tilde\gamma}$:
\a
\sigma &:& \gamma\mapsto{\tilde\gamma} = {\gamma\over(\gamma - 1)}
\label{mapgamma}
\b
A fundamental domain for $\gamma$ is
therefore given by
\a
{\textstyle{1\over 2}} &\leq& \gamma\leq 2
\label{gamfun}
\b
The special point $\gamma =1$
corresponds to the degenerate hierarchy
obtained
from  $t={\textstyle{2\over 3}}$.\par
$\gamma = 2$ corresponds to the symmetric point $t={\textstyle{1\over
2}}$, while $\gamma={\textstyle{1\over 2}}$ is obtained
from the extremal value $t=1$.\par
Clearly different values
of $\gamma$ correspond to hamiltonians having different symmetry
properties: for instance $\gamma =2$
corresponds to the $\sigma$-symmetric hamiltonians, but
such a symmetry is broken for $\gamma\neq 2$.\par
Anyway the equations of motion may have a
bigger symmetry property than the corresponding hamiltonians and a natural
question one can ask is the following: is $\gamma$ a fake or a genuine
parameter in our theory? Stated otherwise, is it possible to redefine
the variables in our theory in such a way that $\gamma$ could be
rescaled to a given fixed value of reference? A simple inspection shows that
this is not the case: it is not possible, under the combined action of
linear transformations for time and space variables and linear mappings of
the fields
$\phi_i(x)\mapsto {\tilde{\phi_i}} = A_{ij} \phi_j$ (with
$det(A_{ij}\neq 0$) to recast our equations (\ref{secondflow})
in a way that $\gamma $ assumes a fixed, specified value.\par
Therefore $\gamma$ is a genuine free parameter in our theory which
labels a continuous class of inequivalent integrable hierarchies.
The integrability of (\ref{sl3unicomp})
is guaranteed for any value of $\gamma$
satisfying (\ref{gamone}) or (\ref{gamtwo}). The restricted (\ref{gamfun})
interval for $\gamma$ corresponds to the class of
$sl(3)$ fundamental hierarchies (the remaining hierarchies
are obtained by $\sigma$-transforming this fundamental class).
\par
$sl(3)$ is the simplest algebra admitting such a structure.
Applying the same considerations here developed to the
$sl(n)$ algebra, we expect that in this case
there exists a continuous class of inequivalent
integrable hierarchies specified by
$n-2$ real parameters.\par
It will be shown later that only the hierarchies corresponding
to the restricted class of values for $\gamma$,
\a
\gamma\geq 1\nonumber
\b
can be supersymmetrically extended while mantaining the integrability
property.

\section{The supersymmetric AKS framework
for the homogeneous grading.}

\indent

In the previous sections we have analyzed the matrix AKS framework
with respect to the homogeneous grading for bosonic hierarchies.\par
In this section I will define and show
the general procedure which allows to extend the AKS construction
to the supersymmetric hierarchies for the homogeneous grading.
\par
This approach furnishes a method to systematically
construct a vast class of super-hierarchies and to automatically
prove their integrability. A manifest $N=1$ supersymmetric formalism
will be used; by no means this implies restriction to $N=1$
super-hierarchies only. It is indeed true (see \cite{{hull},{EH}} for general
considerations and \cite{{ks},{kst}} for an actual construction) that
some of the hierarchies here considered admit an $N=2$
supersymmetry.\par
Before introducing the basic ingredients of such a framework, let me
recall what already stated in the introduction: the matrix AKS framework
for super-hierarchies has already been considered in \cite{IK}, but
only for the principal grading case. The super-hierarchies derived in
such a
case are of super-KdV type and form a rather restricted class
(the hierarchies are put in correspondence with the subclass of
super-Lie algebras which admit a presentation in terms of fermionic
simple roots only). On the contrary, the super-hierarchies derived
within our homogeneous-grading procedure are those of
super-NLS type; apparently they form
a ``wider" class since any bosonic Lie algebra and any super-Lie algebra
can be used to produce their corresponding hierarchies.
More comments on that will be given later.\par
Let us fix at first our conventions concerning the superspace. We denote
with capital
letters the $N=1$ supercoordinate ($X\equiv x, \theta$, with $x$
and $\theta$ real, respectively bosonic and grassmann, variables). \par
The supersymmetric spinor derivative is given by
\a
D \equiv D_X &=& {\partial\over \partial\theta} +\theta
{\partial\over \partial x}
\b
With the above definition $ {D_X}^2 ={\textstyle{\partial\over
\partial x}}$. \par
The supersymmetric delta-function $\Delta (X,Y)$ is a
fermionic
object
\a
\Delta (X,Y) &=& \delta (x-y) (\theta -\eta)
\b
(here $Y\equiv y, \eta $).\par
It satisfies the relations
\a
\Delta (X,Y) &=& -\Delta (Y,X) \quad\quad\quad
D_X\Delta (X,Y) =- D_Y\Delta (X,Y)
\b
Our convention for the
integration over the grassmann variable is
\a
\int d\theta \cdot \theta &=& -1
\b
For any given superfield $F(X)$ we
get then
\a
\int dY \Delta (X, Y )F(Y) &=& F(X)
\b
As in the bosonic case, the
(super)-line integral over a total
derivative gives a vanishing result.\par
At this point it is quite natural to introduce
${\cal L}$ as the
supersymmetrized version of the matrix Lax operator
through the following position:
\a
{\cal L} &=& D_X + \sum_i {\bf \Psi}_i(X) g_i + \Lambda
\label{susylm}
\b
where in the above relation $g_i$'s denote either
the generators of
a semisimple Lie algebra or the generators of a super-Lie algebra;
in the latter case the $g_i$'s can be either bosonic (even elements
${g_i}^{(0)}$),
or fermionic (odd elements ${g_i}^{(1)}$),
see \cite{scheu} for an account on super-Lie algebras.
\par
${\bf \Psi}_i$ denotes the set of $N=1$ currents associated to the
(super)-Lie algebra; they have opposite parity with respect to that of
$g_i$, i.e. the current associated to a bosonic generator
${g_i}^{(0)}$ is fermionic
and conversely to a fermionic generator ${g_i}^{(1)}$
corresponds a bosonic current. In particular the whole set of $N=1$
currents of a standard Lie algebra which admits bosonic generators only
is given by purely fermionic superfields. \par
With the above assumption the second term in the right hand side of
(\ref{susylm}) is fermionic, just like the first term (the spinor
derivative $D$). \par
In the following bosonic and fermionic
superfields will be distinguished by conventionally denoting as
$\Phi_j(X)$
the bosonic superfields and $\Psi_j(X)$ the
fermionic ones.\par
The $N=1$ supercurrents are the generators of the $N=1$
${\hat{\cal G}}$ affinization
of the ${\cal G}$ Lie or super-Lie algebra (see
\cite{div} and \cite{coq} for the affinization of
respectively
bosonic
algebras and superalgebras). This means they satisfy the
following supersymmetric Kac-Moody Poisson brackets algebra
\a
\{ {\bf \Psi}_i (X), {\bf \Psi}_j (Y) \} &=& \sum_k {f^k}_{ij}{\bf\Psi}_k
(Y) \Delta (X,Y) + K_{ij}D_Y\Delta (X,Y)
\label{susyaff}
\b
The above is the $N=1$ extension of the (\ref{aff}) formula. \par
Here ${f^k}_{ij}$ are the (super)structure constants of ${\cal G}$
and $K_{ij} = Str(g_ig_j)$ is the (super)trace in the adjoint
representation of ${\cal G}$. The brackets are either symmetric
or antisymmetric according to the grading of the supercurrents.\par
Just as in the bosonic case the above
algebra will furnish the Poisson brackets structure for the derived
integrable hierarchies.\par
We still need to specify what is $\Lambda$ in (\ref{susylm}). It must be a
constant regular element as in the previously studied case. Here however
an extra-condition appears: due to the fermionic
character of $D$ it seems
unavoidable to assume $\Lambda$ fermionic too in order to
keep a definite statistics for ${\cal L}$; since in the principal
grading case ${\cal L}$ is a sum over a simple-roots set of the ${\cal G}$
(super)algebra, it follows that the above construction works only
if superalgebras are considered which, moreover, admits a presentation
in terms of fermionic simple roots only. This case has been analyzed
in \cite{IK}.\par
When the homogeneous grading is concerned, then $\Lambda$ should assume
the form
\a
\Lambda&\equiv&\lambda H
\b
with $\lambda$ a bosonic spectral parameter and $H$ a generic element in
the Cartan sector of the ${\cal G}$ (super)algebra. \par
Since the Cartan sector is always bosonic, even for super-Lie
algebras, the above consideration seems to
rule out the possibility of introducing integrable hierarchies
in connection with the homogeneous grading. However, supersymmetric
generalizations of the NLS hierarchies have been produced
(\cite{{roe},{brun},{toppan}})
and furthermore it has been shown that at least
some of them satisfy a coset
property \cite{{toppan},{kst}},
which puts them on the same foot as the corresponding bosonic
hierarchies.
It seems therefore rather puzzling that they cannot
be accomodated in an AKS framework.\par
There is however a key point which allows us to overcome the previous
argument: the presence of the $\lambda$ bosonic spectral parameter
makes possible to introduce a sort of ``twisted" statistics
for Laurent series in $\lambda$.
We can indeed assume that the Laurent expansion $B(\lambda)$
is a ``twisted" boson if it is given by an alternating sum of bosonic
and fermionic power series in $\lambda$, such that
\a
B(\lambda) &=& b(\lambda^2) +\lambda \cdot f(\lambda^2)
\label{twb}
\b
with $b(\lambda^2)$, $f(\lambda^2)$ respectively
ordinary bosonic and fermionic Laurent expansions in $\lambda^2$.
\par
Conversely $F(\lambda)$ is a ``twisted" fermion if
\a
F(\lambda) &=& \xi(\lambda^2) +\lambda\cdot \phi(\lambda^2)
\label{twf}
\b
where
$\xi(\lambda^2)$ ($\phi(\lambda^2)$) is an ordinary fermion (boson).\par
``Twisted" bosons and fermions are closed under multiplication with the
same rules as the ordinary bosons and fermions.\par
As already pointed out alternating sums of bosons and fermions already
appeared in the context of GSO projection and the supersymmetric Witten
index.\par
In this respect ${\cal L}$ in (\ref{susylm}) must be considered as a
twisted fermion.\par
The crucial feature in the bosonic AKS picture is the existence
of an {\em uniquely} defined adjoint action which allows diagonalizing
the Lax operator. The same property holds in the supersymmetric case,
but since now we must respect the twisted fermionic character
of ${\cal L}$, the adjoint action should be defined with respect to a
twisted boson.\par
For two generic twisted boson and fermion given by (\ref{twb},\ref{twf})
respectively, we can define the $ad_B (F)$ adjoint action as follows
\a
ad_B(F) &=_{def} [b,\xi]+\lambda^2[f,\phi] +\lambda\cdot
(\{f',\xi\} +[b,\phi])
\label{adjaction}
\b
where the brackets denote the standard commutator and the curly braces
the anticommutator (in consistence with the statistics of the
component fields $b, f,f',\xi,\phi$).
In the above formula
\a
f'&\equiv & t(f)
\label{deft}
\b
is fermionic and $t$
is a linear transformation such that $t^2=1$.\par
This transformation
will be explicitly defined in the next section.\par
The map
\a
F&\mapsto &{\tilde F} = {\tilde \xi} +\lambda\cdot{\tilde \phi}
= ad_B(F)
\b
sends $F$ into a new twisted fermion since ${\tilde\xi},{\tilde\phi}$
\a
{\tilde\xi}&=& [b,\xi]+\lambda^2[f,\phi]\nonumber\\
{\tilde\phi}&=&\{f',\xi\} +[b,\phi]
\b
are respectively bosonic and fermionic.\par
In the following we will need to introduce the $Ad_B (F)$ action
through:
\a
Ad_B(F)&=_{def}& exp(ad_B) (F)=\sum_{n=0}^\infty (ad_B)^n (F)
\b
It turns out there exists an uniquely defined twisted boson $M$
expanded in non-positive powers in $\lambda$, with the boundary
condition $M\rightarrow 0$ for $\lambda\rightarrow\infty$,
\a
M&=& \sum_{k=1}^\infty \lambda^{-2k} ( b_k + \lambda\cdot f_k)
\label{mcompon}
\b
(the $b_k$'s are bosons and the $f_k$'s fermions),
which diagonalizes ${\cal L}$ under its $Ad_M$ action:
\a
{\hat{\cal L}} &=& Ad_M({\cal L})=
\lambda H + D + {\bf \Psi_\alpha}h_{\alpha} +\sum_{k=1}^\infty
\lambda^{-k} R_{k,\alpha}h_{\alpha}
\nonumber\\
&&
\b
$h_\alpha$ are the Cartan generators. For even (odd) values
of $k$, $R_{k,\alpha}$ are fermions (bosons). \par
In particular the $R_{2k,\alpha}$ fermionic quantities provide
the infinite set of hamiltonian densities
for a given hierarchy. We recall that the integration over the
superspace is fermionic, so that
\a
H_{k,\alpha} &=& \int dX R_{2k,\alpha} (X)
\b
are the bosonic hamiltonians, mutually in involution
under the (\ref{susyaff}) Poisson brackets structure.
\par
The compatible flows are defined, for any given superfield $\Xi(X)$, as
\a
{\textstyle{\partial \Xi(X)\over\partial t_{k,\alpha}}} &=& \{
\Xi(X),H_{k,\alpha}\}
\label{susyflows}
\b

\section{The simplest example: the super-NLS hierarchy from the $N=1$
${\hat {sl(2)}}$ algebra.}

\indent

In the previous section the basic ingredients underlining the
supersymmetric AKS framework for the homogeneous grading have been
introduced. Here a more detailed analysis will be given by
illustrating the simplest example of such a construction which arises
from the $N=1$ affinization of the $sl(2)$ algebra. The associated
hierarchy is the super-NonLinear Schr\"{o}dinger hierarchy already
discussed in \cite{toppan}. With respect to that paper the method here
developed can be immediately extended to more complicated
hierarchies.\par
The $sl(2)$ algebra has been introduced in (\ref{sl2algebra}). Its $N=1$
affinization \cite{toppan} is generated
by the $3$ fermionic superfields $\Psi_\epsilon (X)$ (with
$\epsilon\equiv 0,\pm$) and is determined by the following relations:
\a
\{\Psi_0(X),\Psi_0(Y)\}&=&D_Y\Delta(X,Y)\nonumber\\
\{\Psi_0(X),\Psi_\pm(Y)\}&=&\pm 2\Delta(X,Y)\Psi_\pm (Y)\nonumber\\
\{\Psi_+(X),\Psi_-(Y)\}&=&
D_Y\Delta(X,Y)+2\Delta(X,Y)\Psi_0(Y)\equiv{\cal D}_Y\Delta(X,Y)
\label{n1affinsl2}
\b
(any other Poisson bracket is vanishing).\par
The superfields $\Psi_\epsilon(X)$ are decomposed in terms of their
component fields as
\a
\Psi_\epsilon (X) &=& \psi_\epsilon (x) +\theta J_\epsilon (x)
\b
where $\psi_\epsilon$ ($J_\epsilon$) are respectively fermionic
(bosonic). \par
The Poisson brackets for the component fields can be directly
read from (\ref{n1affinsl2}).\par
The upper line in (\ref{n1affinsl2}) specifies the $N=1$ ${\hat{U(1)}}$
Kac-Moody subalgebra generated by $\Psi_0(X)$. In general, as in the
bosonic case, we can introduce charged superfields $V_q(X)$ \\
($q$ is the charge),
assumed to satisfy the following relation with respect to the
$N=1$ ${\hat{U(1)}}$ Kac-Moody generator:
\a
\{\Psi_0(X),V_q(Y)\}&=& q\Delta(X,Y)V_q(Y)
\label{n1charge}
\b
In the above formula $V_q(X)$ can either be a bosonic or a fermionic
superfield. \par
The notion of a (supersymmetric) covariant derivative
${\cal D}$ can be introduced as in the bosonic case through the
following position:
\a
{\cal D}V_q (X)&=& D V_q(X) +q\Psi_0(X)V_q(X)
\label{n1covder}
\b
The covariant derivative is now fermionic and maps a $q$-charged
superfield $V_q(X)$ into a new $q$-charged superfield (of opposite
statistics).\par
It follows immediately from (\ref{n1affinsl2}) that $\Psi_\pm(X)$ are
superfields of definite charge $\pm 2$. For that reason
the right hand side
of the last equation in (\ref{n1affinsl2}) can also be reexpressed by making
use of the fermionic covariant derivative, as shown above.\par
It is worth to mention that the higher order fermionic hamiltonian
densities $R_{2k} (X)$ ($k=1,2,... $) which will be introduced later,
belong to the $N=1$ ${\hat{U(1)}}$ coset, that is they have vanishing
brackets with respect to $\Psi_0(X)$ and the (\ref{n1affinsl2})
Poisson structure:
\a
\{\Psi_0(X),R_{2k}(Y)\}&=&0
\label{n1coset}
\b
It follows that the whole set of $R_{2k}$'s hamiltonian densities is
provided by chargeless differential polinomials in the superfields
${\Psi_\pm(X)}$ and fermionic covariant derivatives acting on them.
The actual demonstration of this property plainly follows the
one already
discussed for the bosonic case and for that reason it will
be omitted here.
\par
At this point we are not yet entitled to plug the above formulas
concerning the $sl(2)$ algebra in the the previous section
construction and derive the associated hierarchy, because we have
still to define the $t$ linear transformation in (\ref{deft}) and
explain its origin.\par
The discussion concerning the role of the $t$-transformation has been
postponed until now in order to introduce at first
the needed algebraic setup,
but it will be immediately clear that no peculiar features of
$sl(2)$ appear in the following reasoning, which admits a
trivial generalizion to any given Lie and super-Lie algebra.\par
It has been already remarked that in the bosonic case, no matter
which Lie algebra and which regular element in its Cartan sector are
chosen, the $\pm$-roots exchange symmetry always allows to perform a
reduction from a two-component fields hierarchy into a single-component
hierarchy. The same feature we wish to preserve in the
supersymmetric case as well.\par
In the particular case of the $sl(2)$ algebra, the
$s_\pm$-roots exchange provides the following automorphisms:
\a
E_+\leftrightarrow E_- \quad\quad&\quad&\quad\quad H\mapsto -H\nonumber\\
\Psi_+(X)\leftrightarrow \Psi_-(X)\quad&\quad&\quad \Psi_0(X)\mapsto
-\Psi_0(X)
\b
The hamiltonian densities of the associated hierarchy turn out to have
a well-defined transformation property with respect to the above $s_\pm$
exchange (i.e. they are eigenfunctions
with eigenvalue $\pm1$),
if the component superfields $\xi_k, \phi_k$ of the twisted
fermionic $F(\lambda)$ series:
\a
F(\lambda) &=& \sum_{k=-\infty}^{+\infty} \lambda^{2k}(\xi_k + \lambda
\cdot \phi_k)
\label{twfdecomp}
\b
satisfy the following transformation properties:
\a
s_\pm: && \xi_k \mapsto (-1)^k\xi_k;\quad\quad\quad \phi_k\mapsto
(-1)^{k+1}\phi_k.
\label{spmtrpr}
\b
Notice in particular that the original Lax operator ${\cal L}$ given by
\a
{\cal L} &=& D + \Psi_0(X)H+\Psi_+(X)E_++\Psi_-(X)E_- + \lambda H
\b
satisfies the above relations for $k=0$.\par
The twisted boson $M(\lambda)$ which diagonalizes ${\cal L}$ can be
decomposed as follows:
\a
M(\lambda) &=& M(\lambda)_> + M(\lambda)_<
\b
where the underscript $\star_>$ ($\star_<$) denotes the projection
over the positive (and respectively negative) roots sector
of any given Lie or super-Lie algebra. In the $sl(2)$ case this is just
the
projection over $E_+$ ($E_-$).
If the bosonic (fermionic) components $b_k$ ($f_k$) in $M(\lambda)$
given in (\ref{mcompon}) are mapped into
\a
s_\pm:&& b_k\mapsto (-1)^k b_k;\quad\quad\quad f_k\mapsto (-1)^k f_k
\label{mparitytra}
\b
then, the adjoint action $ad_M (F)$ as defined in (\ref{adjaction}), sends
$F(\lambda)$ in (\ref{twf}) into a new twisted fermion $ad_M(F)$
whose components satisfy the same $s_\pm$-transformation properties as
(\ref{spmtrpr}), provided that in (\ref{adjaction})
the linear transformation
\a
f(\lambda^2)' &=& t (f(\lambda^2))=_{def} f(\lambda^2)_> -f(\lambda^2)_<
\b
is taken into account.\par
Such a $t$-transformation is therefore necessary in order to mantain a
well-defined transformation property under the
$s_\pm$ symmetry in the supersymmetric case too. As a result the
diagonalized ${\hat{\cal L}}$ Lax operator turns out to be expressed
as
\a
{\hat{\cal L}} &=&
Ad_M ({\cal L})=\lambda H + (D + \Psi_0(X) H) + \sum_{k=1}^\infty
\lambda^{-k} (R_k (X) H)
\b
and, under the $s_\pm$ transformation,
the fermionic hamiltonian densities $R_{2k}(X)$
behave as follows:
\a
s_\pm:&& \quad R_{2k} (X) \mapsto (-1)^{k+1} R_{2k}(X)
\b
Apparently it seems that a certain degree of arbitrariness is involved
in choosing the parity for the transformation properties of the $\xi_k$,
$\phi_k$ components of $F(\lambda)$
in (\ref{twfdecomp}); this is however not true: the parity of
$\xi_k,\phi_k$ (and as a consequence, that of $b_k, f_k$) is uniquely
fixed by the following two requirements:\\
{\em i)} the supersymmetric fermionic hamiltonian densities
are eigenfunctions under the $s_\pm$
transformation
and \\
{\em ii)} the set of first hamiltonian densities
(in the general case denoted as $R_{2k ,\alpha}$ with $k=1$,
just $R_2$
in the specific case of $sl(2)$), should have parity $+1$
(so that $R_2\mapsto R_2$).
\par
This second requirement is due to the fact that the first
hamiltonian densities should guarantee a non-trivial flow
(in the $sl(2)$ case $R_2$ provides chiral equations of motion).
This requirement can be understood also as follows:
it implies the supersymmetric first flow to coincide with the bosonic
first flow when all the fermionic fields are set equal to zero.\par
We already know indeed
that
the first hamiltonian density for the $sl(2)$ hierarchy
can be expressed as
\a
{\cal D} \Psi_+ \cdot \Psi_- + {\cal D}\Psi_- \cdot \Psi_+&&
\b
and has $s_\pm$-parity $+1$.
\par
The corresponding term of same dimension
\a
{\cal D}\Psi_+\cdot \Psi_- - {\cal D}\Psi_- \cdot \Psi_+&&
\b
which has $-1$ parity is a fermionic total derivative and gives trivial
equations of motion.\par
Similarly in the $osp(1|2)$ case we obtain non-trivial equations
of motion if the first hamiltonian density is given by
\a
{\cal D}\Phi_+\cdot \Phi_- - {\cal D}\Phi_-\cdot\Phi_+&&
\b
which has parity $+1$ (here $\Phi_\pm$ are bosonic superfields and
$s_\pm: \Phi_-\mapsto \Phi_+\mapsto -\Phi_-$ as it will be discussed in
the next
section).\par
Here again the $-1$ $s_\pm$-parity term
\a
{\cal D}\Phi_+\Phi_- +{\cal D}\Phi_-\Phi_+&&
\b
is a total derivative.\par
It can be explicitly checked with an iterative proof that at any order
in the expansion over the $\lambda$ spectral parameter,
the  $b_k, f_k$ coefficients of the diagonalizing operator $M(\lambda)$
satisfy the (\ref{mparitytra}) $s_\pm$-transformation properties,
which guarantees the consistency of our procedure.
\par
Therefore the
whole set of algebraic rules is specified to compute the infinite
tower
of hamiltonians, mutually in involution,
for any given Lie or super-Lie algebra.\par
For the specific case of the $sl(2)$ algebra we obtain as $M(\lambda)$
diagonalizing operator, at the lowest orders:
\a
f_{1}&=& {\textstyle {1\over 2}}(\Psi_+E_+ -\Psi_-E_-)\nonumber\\
b_{1}&=& {\textstyle{1\over 4 }}({\cal D}\Psi_+ E_+-{\cal D}\Psi_-
E_-)
\b
The diagonalized Lax operators ${\hat{\cal L}}$ reads
\a
R_1 &=& {\textstyle{1\over 2}} \Psi_+\Psi_-\nonumber\\
R_2&=& {\textstyle{1\over 8}}({\cal D}\Psi_+\cdot \Psi_- + \Psi_+\cdot
{\cal D}
\Psi_-)
\b
The second fermionic hamiltonian density $R_4$ is proportional to
\a
R_4&\propto& {\cal D}^3\Psi_+\cdot\Psi_- -{\Psi_+}{\cal D}^3\Psi_-
\b
The following flows are obtained, in a convenient normalization:
\a
{\textstyle{\partial \Psi_\pm\over \partial t_1}} &=& {\cal D}^2\Psi_\pm
\b
and
\a
{\textstyle{\partial \Psi_\pm\over \partial t_2}} &=&
\pm {\cal D}^4\Psi_\pm \mp 4 \Psi_\pm {\cal D} (\Psi_\mp {\cal D}
\Psi_\pm )
\b
while
for any flow, due to the
above specified coset property of the hamiltonian densities, we get
\a
{\textstyle{\partial \Psi_0\over \partial t_k}} &=& 0
\b
In particular the second flow coincides (apart a normalization factor)
with the two-component super-NLS equation of ref. \cite{toppan}
once set the constraint,
compatible with the equations of motion
\a
\Psi_0&\equiv& 0\nonumber
\b
The single-component superfield  super-NLS equation is recovered by
setting the time being imaginary ($t=-it_2$)
and
\a
\Psi_+={\Psi_-}^\star=\Psi&&\nonumber
\b
The final result is
\a
i{\dot\Psi}&=&D^4\Psi -4\Psi D(\Psi^\star \cdot D\Psi )
\b

\section{The integrable super-hierarchy associated to the $osp(1|2)$
superalgebra.}

\indent

In this section it will be shown that the supersymmetric
AKS framework for the homogeneous grading can be worked out not only for
bosonic Lie algebras (as it is the case for the super-NLS equation), but
also for super-Lie algebras. It will be  analyzed in detail
the simplest example of such
kind of construction, namely the hierarchy derived from the $osp(1|2)$
superalgebra.\par
This superalgebra admits $H$ as a bosonic generator and $F_\pm$ as
fermionic ones and is given by the following relations:
\a
\relax [H, F_\pm]&=& \pm 2 F_\pm\nonumber\\
\{F_+, F_- \} &=& H
\b
Its $N=1$ affinization is realized by the $3$ superfields $\Psi_0 (X)$
(fermionic) and $\Phi_\pm (X)$ (bosonic). Formally it is given by
the same relations as (\ref{n1affinsl2}) with the replacement
$\Psi_\pm \mapsto \Phi_\pm $, but now we have to take into account that
the last Poisson bracket in (\ref{n1affinsl2}) is antisymmetric due to
the bosonic character of $\Phi_\pm$.
\par
The $s_\pm$ algebra automorphism associated with the positive versus
negative roots exchange is now a ${\bf Z}_4$ symmetry. In a chosen
normalization we can define it to be
\a
s_\pm: \quad\quad \quad && H\mapsto -H; \quad F_+\mapsto
-F_-;\quad F_-\mapsto F_+.\nonumber\\
&& \Psi_0\mapsto -\Psi_0;\quad \Phi_+\mapsto -\Phi_-;\quad
\Phi_-\mapsto \Phi_+.
\b
One can easily check that in this case
the same transformation properties
(\ref{spmtrpr},\ref{mparitytra}) under $s_\pm$ for the
twisted fermion $F(\lambda)$, the Lax operator ${\cal L}$ and its
diagonalizing matrix $M(\lambda)$ which hold for bosonic algebras
are verified too.
\par
Moreover here again we find that the hamiltonian fermionic densities
belong to the $N=1$ ${\hat U(1)}$ subalgebra coset generated by $\Psi_0$.\par
We find explicitly, at the lowest orders in the $\lambda$ expansion,
for the diagonalizing matrix $M(\lambda)$:
\a
f_{1} &=& {\textstyle{1\over 2}} (\Phi_+ F_+ - \Phi_-F_-)\nonumber\\
b_{1} &=& {\textstyle{1\over 4}} ({\cal D}\Phi_+F_+-{\cal D}\Phi_-F_-)
\b
while the diagonalized ${\hat {\cal L}}$ Lax operator is given by
\a
{\hat{\cal L}}&=& \lambda H + (D+ \Psi_0 H) +\lambda^{-1}\cdot
{\textstyle
{1\over 2
}} \Phi_+\Phi_- H +\lambda^{-2}\cdot {\textstyle{1\over 8}} (
{\cal D}\Phi_+ \cdot\Phi_- -{\cal D } \Phi_-\cdot\Phi_+)H +
O(\lambda^{-3})\nonumber\\
&&
\b
Up to an overall normalization the first fermionic hamiltonian density
is
\a
R_2 &=& {\cal D} \Phi_+\cdot \Phi_- - {\cal D}\Phi_-\cdot \Phi_+
\b
which has $s_\pm$-parity $+1$.\par
Due to the considerations developed in the previous section, the second
hamiltonian density $R_4$ has parity $-1$.\par
In principle there exists two independent chargeless terms
$X_{1,2}$ having the right dimensions
and
parity $-1$ which can contribute to $R_4$:
\a
X_1 &=& {\cal D}^3 \Phi_+\cdot \Phi_- + {\cal D}^3
\Phi_-\cdot\Phi_+\nonumber\\
X_2 &=& \Phi_+\Phi_-({\cal D}\Phi_+\cdot\Phi_- -{\cal
D}\Phi_-\cdot\Phi_+)
\b
The integrability condition for the hierarchy requires
precisely $R_4\propto X_2$
(notice that the corresponding term associated to the $sl(2)$ algebra
is vanishing due to the fermionic character of $\Psi_\pm$).\par
In a convenient normalization we obtain for the first flow
\a
{\textstyle{\partial \Phi_\pm\over \partial t_1 }} &=& {\cal D}^2
\Phi_\pm \pm 4 \Phi_\pm (\Phi_+\cdot\Phi_-)
\b
and for the second flow
\a
{\textstyle{\partial\Phi_\pm\over\partial t_2}} &=&
{\cal D}( {\cal D} \Phi_\pm\cdot \Phi_+\Phi_-) \pm 2 \Phi_\pm
(\Phi_+\Phi_-)^2
\b
Here again we can consistently set $\Psi_0\equiv 0 $.\par
The single-component superfields hierarchies are recovered by setting
\a
&&\Phi_+=-i{\Phi_-}^\star =\Phi
\b
We get for the first flow the equation
\a
{\dot \Phi} &=&\Phi' +4i \Phi |\Phi|^2
\b
while the second flow, obtained by letting the time imaginary, is
\a
{\dot\Phi}&=& \Phi '|\Phi|^2 -\Phi D\Phi\cdot D\Phi^\star
+2i\Phi|\Phi|^4
\b
(the prime denotes the ordinary spatial derivative).\par
In terms of the component fields we have
\a
\Phi(X) &=& \phi(x) +\theta \psi(x)
\b
with $\phi(x)$ bosonic and $\psi(x)$ fermionic.\par
We get respectively
\a
{\dot \phi} &=& \phi ' +4i\phi|\phi|^2\nonumber\\
{\dot\psi}&=& \psi ' +8i \psi|\phi|^2 + 4i \phi^2\psi^\star
\b
for the first flow and
\a
{\dot\phi}&=&\phi ' |\phi|^2 +2i\phi |\phi|^4 -\phi |\psi|^2\nonumber\\
{\dot\psi}&=& \psi '|\phi|^2 +\phi ' (\psi\phi^\star +2\psi^\star\phi)
-{\phi^\star}' \phi\psi +\nonumber\\
&& 2i\psi|\phi|^4 +4i\phi(|\phi|^2 +\psi\phi^\star +\phi^\star\psi)
 \b
for the second one.\par
Notice in the right hand side of the equation of motion
for the bosonic component the presence of the fermionic field,
which implies a non-trivial coupling.

\section{The $N=1$ ${\hat{sl(3)}}$ hierarchies.}

\indent

In this last section I will construct the $N=1$ supersymmetric
extensions
of the ${\hat{sl(3)}}$ hierarchies introduced in section ($7$).\par
It is a rather unexpected result that integrable supersymmetric
extensions can be produced only for the hierarchies
labelled by the $\gamma$ real parameter belonging to the range
\a
\gamma &\geq & 1
\b
The approach here followed has been already illustrated in detail
in the previous sections, so
that here I will limit myself to furnish
the results.\par
Since in order to reach the above conclusion it
is sufficient to explicitly
compute the first flow only, just this case will be
presented
in this paper. \par
It should be noticed that the
computations in the supersymmetric case are
much more involved than in the bosonic case basically because to get
the $k$-th ordered flow we have to perform a double number
(equal to $2k$)
of diagonalizations. It soon appears that computer is needed to
explicitly obtain even the next simplest flows.
\par
The $N=1$ ${\hat{sl(3)}}$ algebra is generated by the
fermionic superfields
$\Psi_{0,1}, \Psi_{0,2}$ and $\Psi_{\pm i}$ with $i=1,2,3$.\par
The following Poisson brackets are verified
\a
\{\Psi_{0,1}(X),\Psi_{0,1}(Y) \} &=& -2 D_Y\Delta (X,Y)
\nonumber\\
\{\Psi_{0,1} (X),\Psi_{0,2}(Y)\}&=& D_Y\Delta (X,Y)\nonumber\\
\{\Psi_{0,2}(X),\Psi_{0,2}(Y)\}&=& -2D_Y\Delta (X,Y)
\label{ppbb1}
\b
and
\a
\{\Psi_{+i},(X),\Psi_{-i}(Y)\}&=&{\cal D}_Y \Delta (X,Y)
\quad\quad for \quad i=1,2,3.\nonumber\\
\{\Psi_{+1}(X),\Psi_{-3}(Y)\} &=& -\Delta(X,Y)\Psi_{-2}(Y)\quad\quad
\{\Psi_{+2}(X),\Psi_{-3}(Y)\}= \Delta(X,Y)\Psi_{-1}(Y)\nonumber\\
\{\Psi_{+3}(X),\Psi_{-1}(Y)\}&=& -\Delta(X,Y) \Psi_{+2}(Y)\quad\quad
\{\Psi_{+3}(X),\Psi_{-2}(Y)\} = \Delta(X,Y)\Psi_{+1}(Y)\nonumber\\
\{\Psi_{\pm 1} (X), \Psi_{\pm 2}(Y)\} &=& \pm \Delta(X,Y)
\Psi_{\pm 3}(Y)\nonumber\\
&&
\label{ppbb2}
\b
The covariant derivative ${\cal D}$ acts as
\a
{\cal D} \Psi_{\pm 1} &=& D\Psi_{\pm 1} \mp \Psi_{0,1}\Psi_{\pm
1}\nonumber\\
{\cal D} \Psi_{\pm 2}&=& D\Psi_{\pm 2} \mp \Psi_{0,2}
\Psi_{\pm2}\nonumber\\
{\cal D} \Psi_{\pm 3} &=& D \Psi_{\pm 3} \mp (\Psi_{0,1} + \Psi_{0,2}
)\Psi_{\pm 3}
\b
As in section ($7$) the regular Cartan element is chosen to be
\a
\lambda H &=& \lambda ( t H_1 + (1-t) H_2)
\b
with $0\leq t\leq 1$.\par
We will also make use of the variables
\a
\gamma &=& {\textstyle{1\over 3t-1}}\nonumber\\
{\tilde\gamma} &=& {\textstyle{1\over 2-3t}}
\b
The first (bosonic) densities for the diagonalized Lax operator
are
\a
R_{1,1} &=& \Psi_{+3}\Psi_{-3} + \gamma \Psi_{+1}\Psi_{-1}\nonumber\\
R_{1,2} &=& \Psi_{+3}\Psi_{-3} +{\tilde{\gamma}} \Psi_{+2}\Psi_{-2}
\b
which are antisymmetric under the $\pm$-roots exchange
\a
&&\Psi_{+1}\leftrightarrow \Psi_{-1};\quad\quad\Psi_{+2}\leftrightarrow
\Psi_{-2};\quad\quad \Psi_{+3} \leftrightarrow - \Psi_{-3}.
\b
and mutually transforms under the $\sigma$ outer automorphism (see
section
($7$)).\par
The first hamiltonian density $R_{2,1}$ is fermionic. It is given by
\a
R_{2,1} &=& {\textstyle {1\over 2}} \gamma^2({\cal D} \Psi_{+1}\cdot
\Psi_{-1} + {\cal D}\Psi_{-1}\cdot\Psi_{+1})\nonumber\\
&\quad& +{\textstyle{1\over 2}} ( {\cal D}\Psi_{+3}\cdot\Psi_{-3} +
{\cal D}\Psi_{-3}\cdot\Psi_{+3})\nonumber\\
&\quad& +C( \Psi_{+1}\Psi_{+2}\Psi_{-3} -\Psi_{-1}\Psi_{-2}\Psi_{+3})
\label{hamiferdens}
\b
with
\a
C&=& {\textstyle {1\over 12}}(4\gamma{\tilde\gamma}
+5\gamma+4{\tilde{\gamma}} +3)
\label{cvalue}
\b
$R_{2,1}$ is invariant under the $\pm$-roots exchange; the hamiltonian
density $R_{2,2} $ is obtained from the previous expression by replacing
$1\leftrightarrow 2$ and $\gamma\leftrightarrow{\tilde\gamma}$.\par
Notice that, while the relative coefficient of the first two terms in
the right hand side of (\ref{hamiferdens}) is fixed by requiring that
the
correct bosonic limit would be reproduced when setting equal to zero
all the fermionic fields, the coefficient $C$of the third term cannot be
recovered from the bosonic limit. As far as the supersymmetrization
only is
concerned $C$ is a free parameter. However, when the integrability
property
is taken into account,
$C$ must be restricted to be the particular value (\ref{cvalue}).\par
{}From the above hamiltonian, together with the (\ref{ppbb1},\ref{ppbb2})
Poisson
brackets structure, the following set of equations of motion is
derived:
\a
{\dot \Psi}_{\pm 1} &=& \gamma^2 {\cal D}^2\Psi_{\pm 1}
\mp (C{\cal D}\Psi_{\mp 2}\cdot\Psi_{\pm 3} +(1-C)
\Psi_{\mp 2}{\cal D}\Psi_{\pm
3})\nonumber\\
&& \pm \Psi_{\pm 1}((1+C)\Psi_{+3}\Psi_{-3} - C
\Psi_{+2}\Psi_{-2})\nonumber\\
{\dot \Psi}_{\pm 2} &=& \pm ((1-C)\Psi_{\mp 1} {\cal D}\Psi_{\pm 3}
-(\gamma^2 - C) \Psi_{\pm 3} {\cal D}\Psi_{\mp 1})\nonumber\\
&& \pm \Psi_{\pm 2} (( 1+C) \Psi_{+3}\Psi_{-3}
-\gamma^2\Psi_{+1}\Psi_{-1})\nonumber\\
{\dot\Psi}_{\pm 3} &=& \pm ( (C-\gamma^2){\cal D}\Psi_{\pm
1}\cdot\Psi_{\pm2} - C \Psi_{\pm 1}{\cal D}\Psi_{\pm 2})\nonumber\\
&& \pm \Psi_{\pm 3} ( (\gamma^2 + C) \Psi_{+1}\Psi_{-1} +C \Psi_{+2}
\Psi_{-2})
\b
Pewrforming the single-component reduction
\a
\Psi_{+1} &=& \Psi_{-1}\equiv\Psi_1\nonumber\\
\Psi_{+2}&=&\Psi_{-2} \equiv \Psi_2\nonumber\\
\Psi_{+3} &=& -\Psi_{-3}\equiv \Psi_3
\b
and setting $\Psi_{0,1}\equiv\Psi_{0,2}\equiv 0$, we obtain
\a
{\dot \Psi}_1 &=& \gamma^2 {\Psi_1} ' - C D\Psi_2\cdot\Psi_3 -
(1-C)\Psi_2 D\Psi_3\nonumber\\
{\dot\Psi}_2 &=& (1-C)\Psi_1 D\Psi_3 -(\gamma^2 - C) \Psi_3
D\Psi_1\nonumber\\
{\dot\Psi}_3 &=& {\Psi_3}' + (C-\gamma^2) D\Psi_1\cdot\Psi_2 - C \Psi_1
D\Psi_2
\b
In terms of the component fields
\a
\Psi_i (X) &=& \psi_i (x) + \theta \phi_i(x)
\b
with $\psi_i(x)$ fermionic and $\phi_i(x)$ bosonic, we get the following
set of equations
\a
{\dot \psi}_1 &=& \gamma^2 {\psi_1}' - C\phi_2\psi_3 -
(1-C) \psi_2\phi_3\nonumber\\
{\dot\psi}_2 &=& (1-C)\psi_1\phi_3 -(\gamma^2 -
C)\psi_3\phi_1\nonumber\\
{\dot\psi}_3 &=& {\psi_3}' + (C-\gamma^2)\psi_2\phi_1 - C \psi_1\phi_2
\b
for the fermionic components and
\a
{\dot\phi}_1 &=& \gamma^2{\phi_1}'-\phi_2\phi_3 - C {\psi_2}'\psi_3
+(1-C)\psi_2{\psi_3}'\nonumber\\
{\dot\phi}_2 &=& (1-\gamma^2)\phi_1\phi_3 - (1-C)\psi_1{\psi_3}' +
(\gamma^2 - C) \psi_3{\psi_1}'\nonumber\\
{\dot\phi}_3 &=& {\phi_3} ' -\gamma^2\phi_1\phi_2 + C\psi_1{\psi_2}'
+(\gamma^2- C)\psi_2{\psi_1}'
\b
for the bosonic ones.\par
Notice in particular that, when setting equal to zero the fermionic
$\psi_i$ fields, we recover precisely the (\ref{firstflowsusy3})
equations of
motion with $\gamma$ replaced by $\gamma^2$
(and the spatial coordinate
$x$ mapped to $x\mapsto - x$ due to normalization conventions).
Therefore only the
bosonic hierarchies associated to non-negative values of $\gamma$
can be supersymmetrically extended
while preserving integrability. More than that, since the $\sigma$
automorphism sends $\gamma\mapsto{\tilde\gamma}$ (see
(\ref{mapgamma})),
and the corresponding transformed hamiltonian all belong to the
integrable hierarchy, it turns out that only for
\a
\gamma\geq 1\nonumber
\b
we have a supersymmetric integrable extension, which is the result
stated above.

 ~\\~\\
\noindent

{\Large {\bf Conclusions}}

\indent

In this paper an analysis of several aspects of the Lie algebraic
approach towards integrable hierarchies have been furnished. The final
aim consists in arriving at a complete classification of the whole
class of inequivalent hierarchies. \par
Two points have been raised:
the role of the regular element in determining the corresponding
hierarchy and the possible relation between different hierarchies
associated to different gradings.\par
The first problem has been here addressed in the particular case of
the homogeneous-graded hierarchies. It has been shown that indeed
inequivalent hierarchies have been obtained as a consequence and that,
for a
generic Lie algebra, they are labelled by continuous values of some real
parameters.\par
Moreover their mutual transformations under Weyl group or outer
automorphisms actions have been investigated, as well as their coset
property. The possible fields reductions when a symmetry is present
have also been considered.\par
For what concerns the supersymmetric integrable hierarchies, the ``trick"
of introducing alternating series of bosons and fermions in the spectral
parameter expansion, considerably allowed us to enlarge the class of
Lie-algebraic-derived hierarchies, which until now was rather restricted
(associated to a very specific principal-graded construction). As a
simple byproduct of our method
we were able to explicitly produce new supersymmetric
hierarchies, not yet investigated so far.\par
The investigation concerning supersymmetric hierarchies is rather
important if we wish to arrive at a consistent formulation of
discretized
$2$-dimensional gravity. The fact that until now no direct
supersymmetric
matrix model formulation is available, obliges us to bypass this step
and to directly formulate such models in terms of super-${\cal W}$
constraints, associated to some integrable hierarchy, on the partition
function.\par
However there exists a large amount of arbitrariness in performing
such supersymmetrizations and a definite criterium should be found
to extract the ``meaningful" hierarchies.
A very good example of that is one of the features discussed in this
paper, namely that there exists a class of bosonic integrable
hierarchies which seem do not admit an integrable supersymmetric
extension. In the $sl(3)$ case we proved that the standard
procedure to obtain integrable supersymmetric hierarchies
defines supersymmetric extensions only for a restricted class of the
bosonic hierarchies.\par
Another example is associated to KdV (see
\cite{math}): there exists a continuous class of $N=2$ super-KdV
equations, but only for $3$ specific values of the parameter we have
integrability. More than that, only one of these values corresponds to
a nice Lie algebraic setting and the associated hierarchy seems, in some
sort, more fundamental than the others.
\par
It is surprising that such a hierarchy is obtained, through a non-local
Darboux transformation, from the super-NLS hierarchy we have here
discussed. The existence of a Darboux transformation in this very
specific case naturally leads us to ask about the second point
mentioned
in this conclusion, that is: which hierarchies, associated to different
gradings, are truly independent and which, on the contrary,
are related to each other
through a Darboux transformation. This point needs to be clarified
in order to have a complete understanding
of the integrable hierarchy picture and deserves investigation.\par
Besides this major question, let us list some of the topics
which can be easily
addressed in the future: \par
Is it possible to use the same ``trick" of introducing twisted fermionic
and bosonic power series to define supersymmetric integrable hierarchies
for the intermediate grading (i.e. different from the principal and the
homogeneous) case?\par
Next, a rather technical problem: how
to formulate $N=2$ hierarchies in terms of a
manifestly $N=2$ formalism. This is not a fundamental problem because,
as already explained, our constructions accomodates $N=2$ supersymmetric
hierarchies in an $N=1$ manifestly superfield formalism.\par
Furthermore, there exists a certain degree of parallelism between the
integrable
hierarchy formulation on one side and the WZNW reductions (leading to
Toda and coset models) on the other. It is likely that at least some of
the
ideas here discussed can find applications to investigate and
generalize the Witten's
black hole construction.\par
As a final point let me recall that the algebraic machinery here
developed allows, with the help of computer, to explicitly produce
in a systematic way more general hierarchies than those here presented,
associated e.g. to the superalgebra $sl(2|1)$ and so on.
{}~\\~\\

\noindent
{\large{\bf Acknowledgements}}
{}~\\~\\
I am pleased to acknowledge L. Bonora, L. Feher,
E. Ivanov, S. Krivonos, P. Sorba and A. Sorin
for profitable discussions.
{}~\\
{}~\\

\end{document}